\documentclass[aps,
prd,preprintnumbers,
nobibnotes,nofootinbib,floatfix,
amsmath,amssymb,
longbibliography,superscriptaddress
]{revtex4-2}
\usepackage[utf8]{inputenc}
\usepackage{siunitx}
\usepackage{amsfonts}
\usepackage{braket}
\usepackage{mathtools}
\usepackage{cancel}
\usepackage{slashed}
\usepackage{pifont}
\usepackage{soul}
\usepackage{bbm}
\usepackage{bm}
\usepackage{macros_AS}
\usepackage[caption=false]{subfig}
\usepackage[font=small,labelfont=bf]{caption}
\usepackage{yhmath} 

\usepackage{booktabs,array}
\usepackage{hhline}
\usepackage{siunitx}
\usepackage{graphicx}
\usepackage{multirow}

\usepackage{dcolumn}
\usepackage{mathtools}
\usepackage{upgreek}
\newcolumntype{T}{D{.}{.}{10}}
\newcolumntype{E}{D{.}{.}{11}}
\newcolumntype{F}{D{.}{.}{5}}

\usepackage[normalem]{ulem}

\usepackage[dvipsnames]{xcolor}
\usepackage{hyperref}
\hypersetup{
    colorlinks=true,     
    linkcolor=blue,      
    citecolor=blue,      
    filecolor=blue,      
    urlcolor=blue        
}

\makeatletter 
\renewcommand\onecolumngrid{
\do@columngrid{one}{\@ne}%
\def\set@footnotewidth{\onecolumngrid}
\def\footnoterule{\kern-6pt\hrule width 1.5in\kern6pt}%
}
\renewcommand\twocolumngrid{
        \def\footnoterule{
        \dimen@\skip\footins\divide\dimen@\thr@@
        \kern-\dimen@\hrule width.5in\kern\dimen@}
        \do@columngrid{mlt}{\tw@}
}%
\makeatother

\begin{document}
\title{Moments of parton distributions functions of the pion from lattice QCD using gradient flow}
\author{Anthony~Francis}
\affiliation{Institute of Physics, National Yang Ming Chiao Tung University, 30010 Hsinchu, Taiwan}
\author{Patrick~Fritzsch}
\affiliation{School of Mathematics, Trinity College, College Green, Dublin, Ireland}
\author{Rohith~Karur}
\affiliation{Department of Physics, University of California, Berkeley, CA 94720, U.S.A}
\affiliation{Nuclear Science Division, Lawrence Berkeley National Laboratory, Berkeley, CA 94720, USA}
\author{Jangho~Kim}
\affiliation{Lattice Gauge Theory Research Center, Department of Physics and Astronomy, Seoul National University, Seoul 08826, Korea}
\author{Giovanni~Pederiva}
\affiliation{J\"ulich Supercomputing Center, Forschungszentrum Jülich, Wilhelm-Johnen-Straße, 54245 Jülich, Germany}
\affiliation{Center for Advanced Simulation and Analytics (CASA), Forschungszentrum Jülich, Wilhelm-Johnen-Straße, 54245 Jülich, Germany}
\author{Dimitra~A.~Pefkou}
\email{dpefkou@berkeley.edu}
\affiliation{Department of Physics, University of California, Berkeley, CA 94720, U.S.A}
\affiliation{Nuclear Science Division, Lawrence Berkeley National Laboratory, Berkeley, CA 94720, USA}
\author{Antonio~Rago}
\affiliation{$\hbar$QTC \& IMADA, University of Southern Denmark, Campusvej 55, 5230 Odense M, Denmark}
\author{Andrea~Shindler}
\email{shindler@physik.rwth-aachen.de}
\affiliation{Institute for Theoretical Particle Physics and Cosmology, TTK, RWTH Aachen University, Sommerfeldstr. 16, Aachen, 52074, Germany}
\affiliation{Nuclear Science Division, Lawrence Berkeley National Laboratory, Berkeley, CA 94720, USA}
\affiliation{Department of Physics, University of California, Berkeley, CA 94720, U.S.A}
\author{Andr\'e~Walker-Loud}
\affiliation{Nuclear Science Division, Lawrence Berkeley National Laboratory, Berkeley, CA 94720, USA}
\affiliation{Department of Physics, University of California, Berkeley, CA 94720, U.S.A}
\author{Savvas~Zafeiropoulos}
\affiliation{Aix Marseille Univ, Universit\'e de Toulon, CNRS, CPT, Marseille, France}

\begin{abstract}
We present a nonperturbative determination of the pion valence parton distribution function (PDF) moment ratios $\braket{x^{n-1}}/\braket{x}$ up to $n=6$, using the gradient flow in lattice QCD.  
As a testing ground, we employ SU($3$) isosymmetric gauge configurations generated by the OpenLat initiative with a pseudoscalar mass of $m_\pi \simeq 411~\text{MeV}$.  
Our analysis uses four lattice spacings and a nonperturbatively improved action, enabling full control over the continuum extrapolation, and the limit of vanishing flow time, $t\to0$.  
The flowed ratios exhibit O($a^2$) scaling across the ensembles, and the continuum-extrapolated results, matched to the $\MSbar$ scheme at $\mu = 2$ GeV using next-to-next-to-leading order matching coefficients, show only mild residual flow-time dependence. 
The resulting ratios, computed with a relatively small number of configurations, are consistent with phenomenological expectations for the pion's valence distribution, with statistical uncertainties that are competitive with modern global fits.  
These findings demonstrate that the gradient flow provides an efficient and systematically improvable method to access partonic quantities from first principles.  
Future extensions of this work will target lighter pion masses toward the physical point, and applications to nucleon structure such as the proton PDFs and the gluon and sea-quark distributions.
\end{abstract}

\preprint{TTK-25-32}

\maketitle

\section{Introduction}
\label{sec:intro}

Parton distribution functions (PDFs) encode the partonic structure of hadrons and are indispensable for precision QCD phenomenology and searches for new physics at current and future colliders.  
Their accurate determination across the full range of Bjorken-$x$ is a central goal of hadronic physics, and it is particularly critical for the precision program at the Large Hadron Collider (LHC) and its high-luminosity upgrade (HL-LHC), where percent-level control of theoretical systematics will be essential~\cite{Amoroso:2022eow}.

The planned Electron-Ion Collider (EIC) in the United States will provide precise measurements at small-$x$ and over a broad range of $Q^2$, thereby addressing kinematic regions that are currently poorly constrained~\cite{Abir:2023fpo}.  
Likewise, the proposed Large Hadron-electron Collider (LHeC) at CERN would extend deep-inelastic scattering to higher energies and luminosities, offering complementary constraints particularly at very small-$x$ and high-$Q^2$~\cite{LHeC:2020van}.  
These inputs are essential to reduce PDF uncertainties in global analyses, thus enabling both stringent tests of the Standard Model (SM) and reliable searches for physics Beyond the Standard Model (BSM).  
At the same time, progress will remain limited if PDFs can only be constrained indirectly through fits to experimental data.  
A non-perturbative determination of PDFs directly from QCD would provide an independent benchmark, reduce dependence on phenomenological assumptions, and deliver systematically improvable inputs to collider phenomenology.
Lattice QCD, which provides a first-principles formulation of QCD in Euclidean spacetime, is the natural framework for such non-perturbative studies.  
It enables a direct numerical evaluation of hadronic correlation functions and matrix elements, thereby offering a path to investigate hadron structure and parton distribution functions directly from the underlying theory.  

The traditional and long-established method in lattice QCD has been the computation of Mellin moments of PDFs using local twist-2 operators.  
While this approach has provided valuable information for the lowest moments, it faces significant challenges when extended to higher $n$.   
For $n>2$, local twist-2 operators suffer from power-divergent mixing with lower-dimensional operators, so that only special hypercubic irreducible representations (irreps) at $n=3,4$ avoid this issue~\cite{Martinelli:1987zd,Martinelli:1987bh}, albeit with severe signal-to-noise degradation~\cite{Alexandrou:2020gxs,Alexandrou:2021mmi}, while for $n>4$ no safe irreps exist.  

To overcome these limitations, a variety of methods have been proposed to extract information on the $x$-dependence of PDFs from lattice QCD.  
Although they differ in technical implementation, quasi-PDFs, pseudo-PDFs, good lattice cross sections, and related approaches using auxiliary currents or hadronic tensors~\cite{Aglietti:1998mz,Detmold:2005gg,Ji:2013dva,Radyushkin:2017cyf,Ma:2014jla,Braun:2007wv,Chambers:2017dov}, all share a common foundation which is the extraction of PDFs from hadronic matrix elements of bilocal or nonlocal operators that can be computed on the lattice.  
First numerical demonstrations of these approaches, followed by systematic studies and collaborative efforts, have established their practical feasibility~\cite{Lin:2014yra,Lin:2014zya,Chen:2016utp,Lin:2017ani,Alexandrou:2015rja,Alexandrou:2016jqi,Alexandrou:2017huk,Chen:2017mzz,Zhang:2018diq,Alexandrou:2018pbm,Zhang:2018nsy,Alexandrou:2018eet,Lin:2018pvv,Fan:2018dxu,Wang:2019tgg,Lin:2019ocg,Chen:2019lcm,Chai:2020nxw,Lin:2020ssv,Zhang:2020dkn,Li:2020xml,Fan:2020nzz,Gao:2020ito,Zhang:2020rsx,Alexandrou:2020qtt,Gao:2021hxl,Constantinou:2020hdm,Detmold:2021uru,Liu:2017lpe,Liu:2020okp,Ma:2017pxb,Bali:2017gfr,Bali:2018spj,Joo:2020spy,Sufian:2019bol,Sufian:2020vzb,Balitsky:2019krf,Orginos:2017kos,Karpie:2017bzm,Karpie:2018zaz,Karpie:2019eiq,Joo:2019jct,Joo:2019bzr,Radyushkin:2018cvn,Zhang:2018ggy,Izubuchi:2018srq,Bhat:2020ktg,Fan:2020cpa,Sufian:2020wcv,Karthik:2021qwz,HadStruc:2021wmh,Fan:2021bcr,Delmar:2022plq,Salas-Chavira:2021wui,Karpie:2021pap,Gao:2022iex,Xiong:2013bka,Ji:2020ect,Gao:2021dbh,Chen:2020ody,Karpie:2023nyg,Dutrieux:2023zpy,Gao:2023ktu,Gao:2023lny,Gao:2022uhg,Gao:2022ytj,Holligan:2024wpv,Holligan:2024umc,Good:2023ecp,Fan:2022kcb,Liang:2019frk,HadStruc:2021qdf,HadStruc:2024rix,Zimmermann:2024zde}.
These developments show that a direct determination of the $x$-dependence of PDFs from lattice QCD has become a realistic goal.

The central limitation of these approaches lies in the simultaneous requirements of short-distance control and large hadron momenta.  
To suppress higher-twist effects and ensure the validity of the leading-twist expansion, the spatial separation $z$ of the bilocal operator must remain small.  
At the same time, for example in the pseudo-PDF approach, reliable access to the $x$-dependence requires large Ioffe time $\nu = z P_z$, which demands nucleon momenta of several GeV.  
In present-day lattice calculations, $P_z$ typically does not exceed $3$~GeV, restricting the accessible range in $\nu$ and thereby the number of usable data points.  
Achieving higher precision without resorting to model assumptions will require finer lattice spacings to extend this window.  
Moreover, excited-state contamination becomes more severe at large $P_z$, and since high momenta also degrade the statistical signal, both effects introduce additional systematic uncertainties that must be controlled.
These challenges motivate the exploration of alternative strategies.  

In this work we pursue a strategy based on the gradient flow (GF)~\cite{Narayanan:2006rf,Luscher:2010iy,Luscher:2011bx,Luscher:2013cpa}, first introduced in Ref.~\cite{Shindler:2023xpd}, which provides a different approach to the computation of PDF moments and offers a way to bypass some of the obstacles inherent in the standard formulations.  
The method eliminates power-divergent operator mixing, enables controlled continuum extrapolations of higher Mellin moments, and significantly improves the signal-to-noise ratio in the lattice correlators used for their determination.  
In the present study the approach is scrutinized numerically in the non-singlet sector for the pion, where it provides a particularly clean testing ground.  
The method can, however, be naturally extended to singlet combinations, to the nucleon, and more generally to other parton distributions.  
The formulation is general and does not depend on the details of the lattice discretization adopted, making it widely applicable across different lattice QCD setups.
A summary of the results obtained here has been presented in Ref.~\cite{Francis:2025rya}; in the present work, we provide a detailed account of the method, the numerical implementation, and the systematic analyses underlying those results. Preliminary results of this work were presented in Ref.~\cite{Francis:2024koo}.

The remainder of this paper is organized as follows.  
In Sec.~\ref{sec:pdf_moments} we review the operator definitions of Mellin moments and their relation to quantities computable in lattice QCD. 
Sec.~\ref{sec:new_method} presents the new methodology based on the GF and its main theoretical advantages.  
The lattice setup is described in Sec.~\ref{sec:lattice_setup}, while the analysis of systematic effects and results is given in Sec.~\ref{sec:analysis_results}.  
In Sec.~\ref{sec:comparison}, we compare our findings to global fits and other lattice computations, and in Sec.~\ref{sec:reconstruction} we discuss the reconstruction of the full $x$--dependence.  
Finally, Sec.~\ref{sec:conclusions} summarizes our conclusions and outlook.

\section{PDF moments and lattice QCD}
\label{sec:pdf_moments}

For quarks, the Mellin moments of the PDFs are defined as
\be
\langle x^{n-1} \rangle_{q}^{h}(\mu) 
= \int_{0}^{1} dx \, x^{n-1}
\left[ q^{h}(x,\mu) + (-1)^{n}\,\qbar^{h}(x,\mu) \right] \, ,
\label{eq:mellin_moments}
\ee
with $q^{h}(x,\mu)$ and $\qbar^{h}(x,\mu)$ the quark and antiquark distributions of flavor $q$ in hadron $h$, $x$ the Bjorken variable, and $\mu$ the renormalization scale.  
These moments appear naturally in the operator product expansion of deep-inelastic scattering and can be computed on the lattice through matrix elements of local twist-2 operators, thereby providing a direct link between experimental PDFs and lattice QCD.  

For quarks of flavor $q$, the Euclidean twist-2 operators take the form
\be
\widehat{O}^{q}_{\{\mu_1\cdots\mu_n\}}(x)
= O^{q}_{\{\mu_1\cdots\mu_n\}}(x)
- \text{traces} \, ,
\label{eq:twist2_hat}
\ee
with
\be
O^{q}_{\{\mu_1\cdots\mu_n\}}(x)
= \qbar(x)\,\gamma_{\{\mu_1}\,\lrD_{\mu_2}\cdots \lrD_{\mu_n\}}\,q(x) \, .
\label{eq:twist2_operator_quark}
\ee
Here $\lrD_\mu = (\rD_\mu - \lD_\mu)/2$ denotes the symmetrized covariant derivative, with $\rD_\mu$ and $\lD_\mu$ the right- and left-acting derivatives, respectively.  
The curly brackets $\{\cdots\}$ indicate normalized symmetrization of Lorentz indices, while the subtraction of traces over all pairs of indices ensures that the operator transforms irreducibly under O($4$),  the group of rotations and reflections in four-dimensional Euclidean space.

In general, the quark bilinears $\qbar \Gamma q$ contain both flavor-singlet and flavor-non-singlet components.  
In the singlet sector, quark operators such as $\sum_q \widehat{O}^{q}_{\{\mu_1\cdots\mu_n\}}$ mix with the gluonic operators under renormalization.  
In the non--singlet sector, built for example from flavor differences such as $\widehat{O}^{u}_{\{\mu_1\cdots\mu_n\}} - \widehat{O}^{d}_{\{\mu_1\cdots\mu_n\}}$, no such mixing occurs, and the operators renormalize multiplicatively.

In the non-singlet sector, where mixing with gluonic twist-2 operators is absent, the relation between operator matrix elements and PDF moments takes a particularly simple form.  
For a spinless hadron $h(p)$, the forward matrix element reads
\be
\braket{h(p)\,|\,\widehat{O}^{q}_{\{\mu_1\cdots\mu_n\}}\,|\,h(p)}
= 2\,\langle x^{\,n-1}\rangle^{h}_{q}(\mu)\,
\left( p_{\mu_1}\cdots p_{\mu_n} - \text{traces} \right) \, .
\label{eq:twist2_matrixelement}
\ee
This relation makes the connection between local operators and Mellin moments transparent, but its practical use in lattice QCD is severely complicated by renormalization issues.  
For $n>2$, twist-2 operators mix with lower-dimensional operators, and the corresponding coefficients diverge as powers of the lattice spacing $a$.  
Such power-divergent mixing obstructs a controlled continuum limit.  
For $n=3$ and $n=4$, one can construct operators belonging to special irreducible representations of the hypercubic group that are protected from this mixing~\cite{Martinelli:1987zd,Martinelli:1987bh}.  
However, only operators with all Lorentz indices different fall into these safe irreps, which forces the use of hadron states carrying nonzero spatial momentum in multiple directions in Eq.~\eqref{eq:twist2_matrixelement}.  
This requirement leads to a severe deterioration of the signal-to-noise ratio, as demonstrated in recent numerical studies~\cite{Alexandrou:2020gxs,Alexandrou:2021mmi}.  
For $n>4$, no suitable irreps exist, and the determination of higher Mellin moments using local twist-2 operators has therefore remained beyond reach.

In relating hadronic matrix elements to Mellin moments, one must account for their connection to the valence and sea sectors, which needs to be properly understood.   
For odd values of $n$, twist-2 operators project directly onto valence moments, and in the SU($2$) isospin limit disconnected diagrams cancel in non-singlet combinations, so that only connected insertions need to be computed.  
For even $n$, the corresponding operators yield quark-plus-antiquark (singlet) moments, but in the SU($3$) symmetric limit the disconnected diagrams cancel in octet combinations, and for the pion the connected $u$-insertion alone reproduces twice the valence moment. This is particularly  relevant in this work, as the computations are performed in the SU($3$) flavor-symmetric limit.
In the next section we introduce the GF method, which provides a different approach to the direct computation of PDF moments and offers a way to bypass some of the obstacles inherent in the standard approach.

\section{Flowed moments}
\label{sec:new_method}

The GF~\cite{Narayanan:2006rf,Luscher:2010iy,Luscher:2011bx,Luscher:2013cpa} provides a systematic framework to overcome the renormalization challenges of local twist-2 operators.  
The basic idea, first proposed in Ref.~\cite{Shindler:2023xpd}, is that the flow acts as an intermediate regulator: gauge and fermion fields are evolved to positive flow time $t$, which smooths ultraviolet fluctuations over a radius of order $\sqrt{8t}$ and suppresses short-distance singularities of composite operators.  
This procedure allows one to recover the correct space-time symmetries after the continuum limit is taken, and the physical PDF moments are then obtained through a short flow-time expansion (SFTX)~\cite{Luscher:2011bx,Suzuki:2013gza,Luscher:2013vga} combined with a matching procedure performed in the continuum theory.

Besides the usual renormalization of the bare parameters of QCD, correlation functions built from flowed fields at $t>0$ do not require further renormalization for gauge fields.  
Flowed fermion fields, however, must be multiplicatively renormalized,
\be
\chi_R(t,x) = Z_\chi^{1/2}\, \chi(t,x), 
\qquad 
\chibar_R(t,x) = Z_\chi^{1/2}\, \chibar(t,x),
\label{eq:chi_ren}
\ee
with $Z_\chi$ the flowed fermion field renormalization constant.  
Unlike standard local operators at $t=0$, which in general require operator-specific renormalization factors and suffer from mixings with operators of equal or even lower dimension (leading to power divergences), any local operator constructed from flowed fermion fields renormalizes simply with the appropriate power of $Z_\chi^{1/2}$.  
For example,
\be
\mcO_\Gamma(t,x) = \chibar(t,x)\, \Gamma(t,x)\, \chi(t,x) \, ,
\label{eq:bilinear_flowed}
\ee
renormalizes multiplicatively with a single factor of $Z_\chi$, where $\Gamma(t,x)$ denotes a generic matrix in color and Dirac space, which may also include flowed gauge fields, as is the case for twist-2 operators.  

For practical lattice applications it is convenient to define $Z_\chi$ in a regularization-independent scheme, so that renormalization and matching can be carried out consistently.  
A particularly useful choice is the so-called \emph{ringed fields}~\cite{Makino:2014taa} ($\rchi$ and $\rchibar$), defined through the SU($N_c$)-gauge-invariant condition
\be 
\left\langle \rchibar_r(x,t)\,{\overset\leftrightarrow{\slashed{D}}}\,\rchi_r(x,t) \right\rangle
= - \frac{N_c}{(4 \pi)^2 t^2}\,,
\label{eq:ringed_condition}
\ee
which is imposed separately for each fermion flavor $r$ in the flowed operator.  
In dimensional regularization, the definition of ringed fields leads to a finite relation with renormalized fields through a normalization factor $\zeta_\chi$.  
This factor has been computed in the $\MS$ scheme at one loop~\cite{Makino:2014taa}, extended to two loops in Refs.~\cite{Harlander:2018zpi,Artz:2019bpr}, and can be straightforwardly converted to the $\MSbar$ scheme.  
Composite operators built from ringed fermions are automatically finite, since with ringed fields the overall factor $Z_\chi$ cancels, simplifying the continuum limit.  

After taking the continuum limit of hadronic matrix elements at fixed flow time $t>0$, their connection to renormalized operators at $t=0$ is established through the SFTX~\cite{Luscher:2011bx,Suzuki:2013gza,Luscher:2013vga}.  
Using the ringed fields defined above, we introduce the flowed traceless twist-2 operators as  
\be 
\widehat{\rmcO}_n^{rs}(t,x) = 
\rmcO_n^{rs}(t,x) - \text{trace terms involving }\delta_{\mu_i \mu_j}\,,
\label{eq:flowed_t2}
\ee 
with 
\be 
\rmcO_n^{rs}(t,x) = \rchibar^r(x,t)\, \gmuopen1 \lrDmu2 \cdots \lrDmuclose{n}\, \rchi^s(x,t)\,,
\label{eq:flowed_On}
\ee 
so that all traces vanish explicitly, and where the covariant derivatives contain flowed gauge fields.
For simplicity, we restrict the notation to $r \ne s$, corresponding to a specific flavor non-singlet component. The same construction applies to any non-singlet combination, including those used in this work.

The SFTX of these operators takes the form
\be 
\widehat{\rmcO}_n^{rs}(t) = \zeta_n(t,\mu)\,\widehat{O}_{n}^{rs}(\mu) + O(t) \,,
\label{eq:sftx_match}
\ee
where $\mu$ is the renormalization scale.  
The operator $\widehat{O}_{n}^{rs}(\mu)$ is renormalized in the same scheme used to compute the matching coefficients $\zeta_n(t,\mu)$, which themselves depend on the UV renormalization prescription for the flowed fermion fields (here: ringed fields).  
For the operators $\rmcO_n(t,x)$ in Eq.~\eqref{eq:flowed_On}, the SFTX can in principle contain power divergences of the type $1/t^m\, \delta_{\mu_i \mu_j}$, where $m=(d_n-d_{n'})/2$ depends on the operator dimension $d_n$ and that of lower-dimensional operators $d_{n'}$.  
Once O($4$) symmetry is restored, however, these terms are organized accordingly, and the SFTX of symmetrized and traceless operators $\widehat{O}_n$ receives contributions only from the corresponding traceless operators at $t=0$~\cite{Shindler:2023xpd}.  
The coefficients $\zeta_n(t,\mu)$ are obtained by matching off-shell amputated one-particle-irreducible (1PI) Green's functions with flowed and unflowed operators.  
In perturbation theory they take the form
\be 
\zeta_n(t,\mu) = 1+ \sum_{k=1}^\infty 
\left(\frac{\gbar^2(\mu)}{(4\pi)^2}\right)^k \zeta_n^{(k)}(t,\mu)\,,
\label{eq:zeta_pert}
\ee
where the coefficients are independent of soft scales, and $\gbar(\mu)$ is the strong coupling renormalized at a scale $\mu$.  
The next-to-leading order (NLO) calculation was presented in Ref.~\cite{Shindler:2023xpd}, and has since been extended to $O(\gbar^4)$~\cite{inpreparation}.  

The form of Eq.~\eqref{eq:sftx_match} shows that flowed matrix elements can be related to physical twist-2 operators in a purely multiplicative way, thereby opening the door to the computation of PDF moments of arbitrary order.  
In practice, one constructs the symmetrized and traceless operator $\widehat{\rmcO}_n^{rs}(t)$, computes its matrix elements from three-point functions, and ensures that the physical separation between the interpolating fields and the flowed operator insertion is larger than the flow radius $\sqrt{8t}$.  
The relation between the flowed matrix elements and the moments of the PDF is then identical to that at $t=0$, as given in Eq.~\eqref{eq:twist2_matrixelement}.
After the continuum limit is taken, the renormalized moment is obtained as
\be 
\left\langle x^{n-1} \right\rangle (\mu) 
= \zeta_n^{-1}(t,\mu)\, \left\langle x^{n-1} \right\rangle(t) \,,
\label{eq:x_MS}
\ee
with residual flow-time dependence arising from O($t$) contributions of higher-dimensional operators and from the truncation of the perturbative expansion in $\gbar^2$.
These systematics are analyzed in Sec.~\ref{ssec:t0_extrapolation}, where we study the extrapolation of flowed moments to $t\to 0$ in the continuum.

Another important advantage of the GF is the improvement in the statistical signal of lattice correlation functions.  
First, the flow itself provides a smoothing of ultraviolet fluctuations over a radius $\sqrt{8t}$, which naturally reduces statistical noise in composite operators.  
Second, since the matching coefficients $\zeta_n(t,\mu)$ are independent of the specific choice of Lorentz indices, one is free to select all indices in the temporal direction.  
This avoids the need to inject nonzero spatial momentum, which in conventional approaches significantly degrades the signal-to-noise ratio, especially for higher moments.  

In summary, the flowed moments method provides a multiplicative renormalization for all twist-2 operators, independent of $n$, and enables a controlled continuum limit for any moment without the need for power-divergent subtractions in the lattice spacing.  
The associated matching coefficients $\zeta_n(t)$ can be computed perturbatively, and the statistical quality of the relevant correlators is significantly improved thanks to both the smoothing of ultraviolet fluctuations and the possibility of working with purely temporal indices.  
Together, these features make flowed moments particularly clean lattice quantities and a powerful and general framework for the numerical determination of PDF moments in lattice QCD.
As we will discuss in Sec.~\ref{ssec:continuum_limit}, the method also offers additional advantages in terms of cutoff effects, which further strengthen its suitability for precision studies.  

\section{Lattice QCD setup}
\label{sec:lattice_setup}

\begin{table*}[t]
\begin{ruledtabular}
\begin{tabular}{c|cccccccccc}
label & $a~[\text{fm}]$ & $t_0/a^2$ & $\beta$ & $\kappa_{ud}=\kappa_s$ & $T/a \times (L/a)^3$ & $N_{\text{cfg}}$ & $\tau_s/a$ & $t_{\text{min}}/a^2$ & $t_{\text{max}}/a^2$ & $\delta t/a^2$ \\
\hline
{\tt a12m412\_mL6.0} ({\tt a12})  & 0.12  & 1.4868(04) & 3.685 & 0.1394305 & $96\times24^3$ & 838 & 35,40 & 0.1 & 4.3  & 0.1 \\
{\tt a094m412\_mL6.2} ({\tt a094})& 0.094 & 2.4400(01) & 3.8   & 0.138963  & $96\times32^3$ & 700 & 35,40 & 0.2 & 6.8  & 0.2 \\
{\tt a077m412\_mL7.7} ({\tt a077})& 0.077 & 3.6239(12) & 3.9   & 0.138603  & $96\times48^3$ & 400 & 40,42 & 0.3 & 9.9  & 0.3 \\
{\tt a064m412\_mL6.4} ({\tt a064}) & 0.064 & 5.2471(26) & 4.0   & 0.138272  & $96\times48^3$ & 500 & 40,42 & 0.5 & 13.5 & 0.5 \\
\end{tabular}
\end{ruledtabular}
\caption{
Parameters and measurement setup of the four gauge ensembles used in this work.  
The first six columns list the ensemble parameters: lattice spacing $a$, Wilson-flow scale $t_0/a^2$, gauge coupling $\beta$, hopping parameter $\kappa_{ud}=\kappa_s$, lattice volume $T/a\times(L/a)^3$, and number of independent analyzed configurations $N_{\text{cfg}}$.  
The remaining columns summarize the measurement setup: source-sink separations $\tau_s$, and the minimum, maximum, and spacing of the flow times $t/a^2$ used in the analysis.
}
\label{tab:ensembles}
\end{table*}

The four ensembles employed in this study were generated by the OpenLat initiative~\cite{Cuteri:2022erk,Cuteri:2022oms,Francis:2022hyr,Francis:2023gcm} within the Stabilized Wilson Fermion (SWF) framework~\cite{Francis:2019muy}, using a Lüscher-Weisz improved gauge action~\cite{Luscher:2011bx,Curci:1983an} and an exponentiated, non-perturbatively improved clover action (expClover) with $N_f=3$ degenerate quark flavors. The bare parameters are tuned to produce a pseudoscalar mass $m_\pi \simeq 411~\text{MeV}$. The ensemble labels and parameters are summarized in Table~\ref{tab:ensembles}.

Because published results~\cite{Brommel:2005ee,Brommel:2007zz,Bali:2013gya,Abdel-Rehim:2015owa,Oehm:2018jvm,Loffler:2021afv} display incompatible $m_\pi$ trends for the moments and the ratios and no single chiral form is clearly preferred, we refrain from making any quantitative statement about the expected corrections toward the physical point.
The observed discrepancies in the lattice results underscores the need for a systematic and robust study of the pion-mass dependence in future work.

The scale is set via the GF parameter $t_0/a^2$~\cite{Luscher:2010iy}, 
with the values for the four ensembles listed in Table~\ref{tab:ensembles}. 
To determine $t_0/a^2$, we compute $t^2\langle E(t)\rangle$, where the action density $E(t)$ is defined using the clover discretization of the field-strength tensor.
The GF evolution is implemented with the Wilson gauge action, using a fixed flow-time step of $t_\epsilon = 0.04$, and the value of $t_0/a^2$ is obtained by quadratic interpolation of $t^2\langle E(t)\rangle$.
A detailed description of the computation and determination of $t_0/a^2$ will be provided in a forthcoming publication.
Converting to physical units using $\sqrt{t_0}=0.14474(57)\,\text{fm}$~\cite{FlavourLatticeAveragingGroupFLAG:2024oxs,RBC:2014ntl,RQCD:2022xux,Bruno:2016plf,BMW:2012hcm} gives approximate lattice spacings $a \simeq [0.12,\,0.094,\,0.077,\,0.064]~\text{fm}$.
In the remainder of this section we define the three-point correlation functions used to extract the PDF moments and detail our numerical strategy for their computation on these ensembles.

\subsection{Three-point correlation functions}
\label{ssec:3pfunc}

To extract the flavor non-singlet flowed moments from lattice QCD, we compute connected fermionic three-point correlation functions of the form
\be
C_{\mu_1\cdots\mu_n}^{uu}(t;\tau_s,\tau_{\mcO}) 
= a^6 \sum_{\bx,\by}\,
\braket{P^{du}(\bx,\tau_s)\, \mcO^{uu}_{\mu_1\cdots\mu_n}(t;\by,\tau_{\mcO})\, P^{ud}(0)}_c \,,
\label{eq:3ptfull}
\ee
where $t$ denotes the flow time of the operator, 
$y=(\by,\tau_{\mathcal{O}})$ its location, and $\tau_s$ the sink time.  
Here $P^{ud}=\overline{\psi}^u \gamma_5 \psi^d$ is an interpolating field with the quantum numbers of the $\pi^+$.  
Since we work with $N_f=3$ degenerate quarks, the choice of flavor is immaterial, though explicit labels are kept for bookkeeping.  
The source is placed at the origin for simplicity, and the spatial sums over $\bx$ and $\by$ project both the sink and the operator insertion to zero spatial momentum.  

The operators $\mcO^{qq}_{\mu_1\cdots\mu_n}$ are defined in terms of the discretized symmetric covariant derivative,  
\begin{align}
\lrD_\mu &= \tfrac{1}{2}\left(\rD_\mu-\lD_\mu\right),
\label{eq:covder}\\
\rD_\mu\chi^q(t,x) &= \tfrac{1}{2a}\Big[V_{\mu}(t,x)\chi^q(t,x+a\hat{\mu})-V_{\mu}^{\dagger}(t,x-a\hat{\mu})\chi^q(t,x-a\hat{\mu})\Big],
\label{eq:covderfwd}\\
\chibar^q(t,x)\lD_\mu &= \tfrac{1}{2a}\Big[\chibar^q(t,x+a\hat{\mu})V_{\mu}^{\dagger}(t,x)-\chibar^q(t,x-a\hat{\mu})V_{\mu}(t,x-a\hat{\mu})\Big]\,,
\label{eq:covderbwd}
\end{align}
where $V_\mu(t,x)$ and $\chi^q(t,x)$ ($\chibar^q(t,x)$) are the flowed gauge and fermion (antifermion) fields, respectively, evolved according to the GF equations discretized following Refs.~\cite{Luscher:2010iy,Luscher:2013cpa}. 
 
After Wick contractions, and retaining only the connected part, Eq.~\eqref{eq:3ptfull} becomes
\be
C^{uu}_{\mu_1\cdots\mu_n}(t;\tau_{\mathcal{O}},\tau_s) 
= -a^3 \sum_{\by} \Big\langle 
\tr\Big\{ \Sigma_{ud}^\dagger(t;0,y;\tau_s)\,\gamma_5\,
\Gamma_{\mu_1\cdots\mu_n}(t,y)\, S_u(t;y,0) \Big\}
\Big\rangle ,
\label{eq:3pt_flowed}
\ee
where the operator insertion is encoded in
\be
\Gamma_{\mu_1\cdots\mu_n}(t,y) 
= \gamma_{\mu_1}\,\lrD_{\mu_2}(t,y)\cdots\lrD_{\mu_n}(t,y) \,,
\label{eq:gamma_insert}
\ee 
and the flowed forward and sequential propagators,
$S_u(t;y,0)$ and $\Sigma_{ud}(t;y,0;\tau_s)$, are given by
\be
S_u(t;y,0) = a^4 \sum_v K(t;y,v)\, S_u(t=0;v,0)\,,
\label{eq:flowed_prop}
\ee
\be
\Sigma_{ud}(t;y,0;\tau_s) = a^4 \sum_v K(t;y,v)\, \Sigma_{ud}(t=0;y,0;\tau_s)\,,
\label{eq:flowed_seq}
\ee  
with $K$ the kernel of the discretized GF differential operator.  

We first determine the unflowed sequential propagator using the sequential source-through-the-sink method~\cite{Bernard:1985ss}.  
After computing the forward propagator $S_u(t=0;y,0)$, the sequential propagator is defined as  
\be
\Sigma_{ud}(t=0;y,0;\tau_s) 
= a^3 \sum_{\bx} S_u(t=0;y,x_s)\,\gamma_5\, S_d(t=0;x_s,0)\,,
\label{eq:seq_def}
\ee
and is obtained by solving the lattice Dirac equation with $D_u$ the Dirac operator for flavor $u$ and the source given by $\gamma_5\, S_d(t=0;x,0)$ located at $x_4=\tau_s$. 
The flowed propagators in Eqs.~\eqref{eq:flowed_prop} and \eqref{eq:flowed_seq} are then obtained by solving the GF equation with the unflowed propagators, $S_u(t=0;y,0)$ and $\Sigma_{ud}(t=0;y,0;\tau_s)$, as initial conditions.
In this setup, one forward propagator inversion and one sequential inversion per sink time $\tau_s$ are required. 

\subsection{Contraction optimization}
\label{ssec:contractions}

The computation of three-point functions with multiple covariant derivatives quickly becomes costly, as each discretized derivative introduces several terms.  
In this subsection we describe an algorithmic strategy to minimize the number of required contractions by exploiting symmetries of the lattice discretization and by reusing intermediate results.

For a given operator $\mcO^{qq}_{\mu_1...\mu_n}$, before symmetrization and trace subtraction over indices, the three-point function contains a total of $4^{n-1}$ terms, corresponding to the four contributions from each discretized symmetric covariant derivative, as seen from Eq.~\eqref{eq:covder}. 
The total number of contractions can be reduced by noting that the terms contained in $\cdots\lD_{\mu_i}\cdots$ are identical to those in $\cdots\rD_{\mu_i}\cdots$,  
except that terms including $V_{\mu_i}$ are shifted by $+a \hat{\mu}_i$, while those with $V^{\dagger}_{\mu_i}$ are shifted by $-a \hat{\mu}_i$.  
This correspondence follows directly from comparing the forward derivative in Eq.~\eqref{eq:covderfwd} with the backward derivative in Eq.~\eqref{eq:covderbwd}.  
As a result, one can reduce the contractions from $4^{n-1}$ to $2^{n-1}$ by first computing the contracted terms contained in the forward-derivative-only contribution, and then reconstructing the full symmetrized derivative contribution by shifting and linearly combining them with the appropriate coefficients.  
\begin{table}[]
  \begin{center}
  \begin{tabular}{c|c|ccccccc}
  $n$ & $\widehat{O}_{n,4...4}$ &&&&&&& \\
  \hline
  2& $\widehat{O}_{44}$&$\sum_{i}O_{ii}$ &$O_{44}$ &&&&& \\
  && -1/3 & 1 &&&&&\\
  \hline
  3& $\widehat{O}_{444}$& $\sum_{i}O_{\{ii4\}}$ & $O_{444}$ &&&&& \\
  && -1 & 1 &&&&& \\
  \hline
  4& $\widehat{O}_{4444}$& $\sum_{i}O_{iiii}$ & $\sum_{i,j>i}O_{\{iijj\}}$ & $\sum_{i}O_{\{ii44\}}$ & $O_{4444}$ &&&\\
  && 1/5 & 2/5 & -2 & 1 &&& \\
  \hline
  5& $\widehat{O}_{44444}$&$\sum_{i}O_{\{iiii4\}}$ & $\sum_{i,j>i}O_{\{iijj4\}}$ & $\sum_i O_{\{ii444\}}$ & $O_{44444}$ \\
  &&1 & 2 & -10/3 & 1 &&&\\
  \hline
  6& $\widehat{O}_{444444}$&$\sum_i O_{iiiiii}$ & $\sum_{i,j>i}O_{\{iiiijj\}}$ & $O_{\{112233\}}$ & $\sum_i O_{\{iiii44\}}$ & $\sum_{i,j>i}O_{\{iijj44\}}$ & $\sum_i O_{\{ii4444\}}$ & $O_{444444}$ \\
  &&-1/7 & -3/7 & -6/7 & 3     & 6 & -5 & 1 \\
  \end{tabular}
  \caption{
Twist-2 operator bases used in this work.  
Each row shows the explicit decomposition of $\widehat{O}_{n,4\ldots4}$ into a linear combination of operators listed to its right, with the corresponding coefficients given in the row below.  
For example, for $n=2$, $\widehat{O}_{44} = -\tfrac{1}{3}\sum_i O_{ii} + O_{44}$, with $i \in \{1,2,3\}$.  
Braces $\{i_1\ldots i_n\}$ denote normalized symmetrization over indices.
}
  \label{tab:tracelessnorm}
  \end{center}
\end{table}

We start by defining the binary vector $\underline{l} = (l_2,\ldots,l_n) \in \{0,1\}^{n-1}$.  
The specific expression for Eq.~\eqref{eq:3pt_flowed} can then be efficiently evaluated as
\be \label{eq:Cefficient}
C_{\mu_1...\mu_n}^{uu}(t;\tau_s,\tau_{\mathcal{O}}) = 
\frac{1}{2^{n-1}} \sum_{\underline{l}\in\{0,1\}^{n-1}} \frac{1}{(2a)^{n_4}}
\sum_{k=0}^{n_4} \binom{n_4}{k}\,
\mathcal{C}^{uu}_{\mu_1\mu_2...\mu_n}\Big(t;\tau_s,\tau_{\mathcal{O}}-\Delta \tau_n(\underline{l})+k;\underline{l}\Big) , 
\ee
where $n_4$ is the number of temporal covariant derivatives among the $n-1$ defining the operator, and we define
\be
\mathcal{C}^{uu}_{\mu_1\mu_2...\mu_n}(t;\tau_s,\tau_{\mathcal{O}};\underline{l}) 
= -a^3 \sum_{\by}\Big\langle\tr\big\{ \Sigma^{\dagger}_{ud}(t;0,y;\tau_s)\gamma_5\gamma_{\mu_1}\,
\mathcal{P}^{\ell_2}_{\mu_2}\mathcal{P}^{\ell_3}_{\mu_3}\cdots\mathcal{P}^{\ell_{n}}_{\mu_{n}}S_{u}(t;y,0)\big\}\Big\rangle .
\ee
The forward- and backward-shifting operators acting on a generic field $\phi$ are defined as
\bea
  \mathcal{P}^{1}_{\mu}\phi(x) &\equiv& V_{\mu}(t,x)\,\phi(x+a\hat{\mu}) \;, \\
  \mathcal{P}^{0}_{\mu}\phi(x) &\equiv& -V^{\dagger}_{\mu}(t,x-a\hat{\mu})\,\phi(x-a\hat{\mu}) \;,
  \label{eq:shift_op}
\eea 
while $\Delta \tau_n$ denotes the time shift, relevant only for temporal derivatives, of each term in the linear combination:
\be
\Delta \tau_n(\underline{l}) = \sum_{\substack{2 \le i \le n \\ \mu_i = 4}} l_i\,.
\ee 

For derivatives with spatial indices $\mu_i\in\{1,2,3\}$, the definite three-momentum projection implies that the terms stemming from the forward- and backward-derivative in the operator are numerically identical up to an overall phase, which is unity for the unboosted case considered here.  
Consequently, the total number of contractions required for each three-point function $C^{uu}_{\mu_1...\mu_n}(t;\tau_s,\tau_{\mathcal{O}})$ is reduced from the naive $4^{n-1}$ to $2^{n_4}$.

This procedure must be carried out once for each unique Lorentz index combination $\mu_1 \ldots \mu_n$ required to form the traceless operators $\widehat{\mcO}_{n,4 \ldots 4}$.  
The total number of such combinations is $4$, $10$, $40$, $136$, and $544$ for $n=2,3,4,5,$ and $6$, respectively, with the explicit linear combinations given in Table~\ref{tab:tracelessnorm}.  
To further reduce computational cost, we avoid measuring the operators of each $n$ iteratively; instead, they are organized ``depth first", allowing intermediate expressions to be reused. For instance, $\mcP^1_1 S$ can be recycled in the evaluation of $\mcP^1_2 \mcP^1_1 S$.  

\begin{figure}[t]
    \centering
    \includegraphics[width=0.99\linewidth]{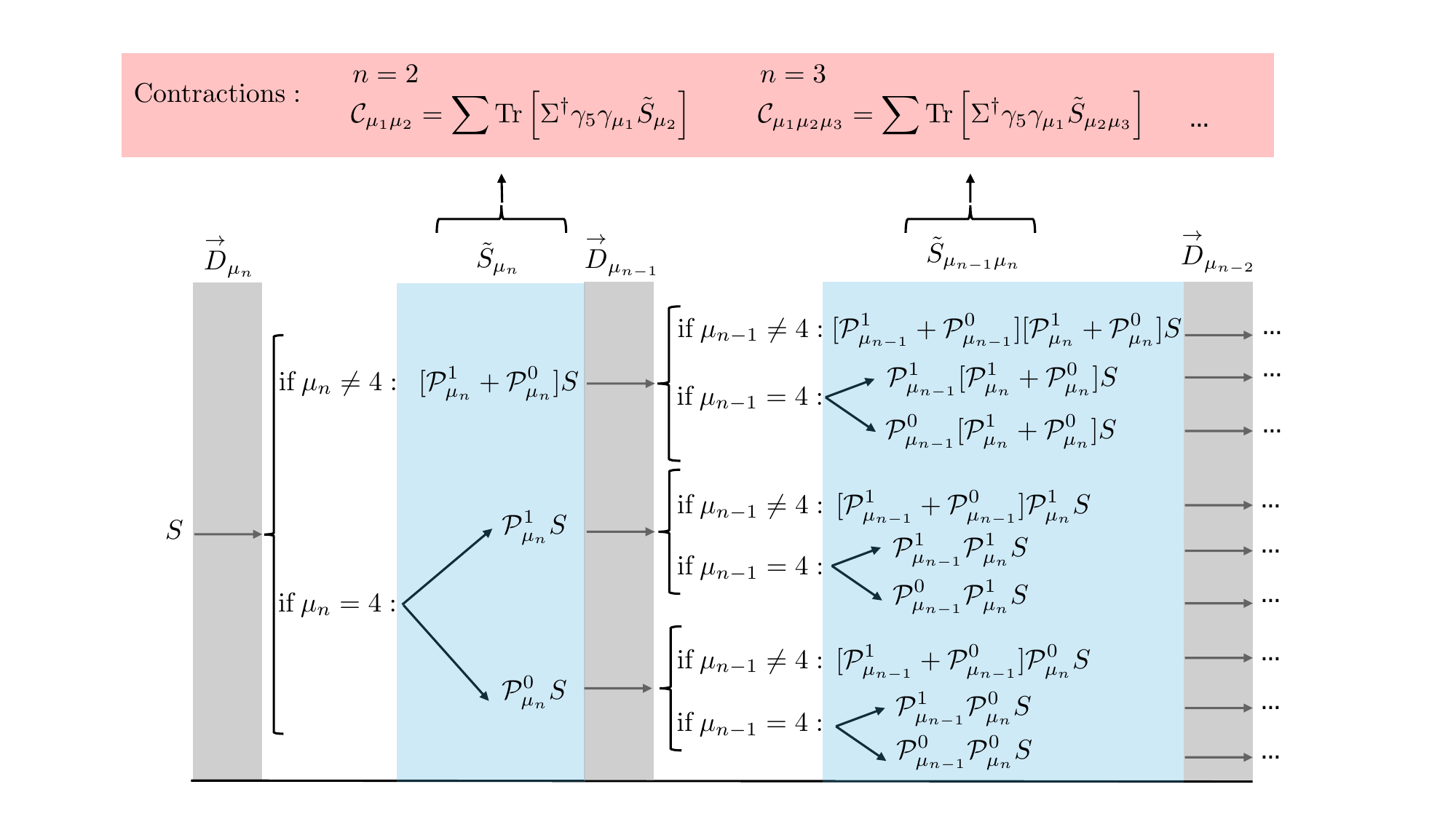}
    \caption{Schematic representation of the optimized workflow for the contractions necessary to compute the three-point functions. This is repeated at each $\tau_s$ and $t$, while the arguments of the propagators and other various objects defined in the text are left implicit. We start with the forward propagator $S$ and act on it with the forward covariant derivative
    $\protect\overrightarrow{D}_{\mu_n}$. For temporal $\mu_n$, we separately save $\tilde{S}_{\mu_n} = \{\mathcal{P}^1_{\mu_n} S, \mathcal{P}^0_{\mu_n} S\}$, while for spatial $\mu_n$, we save the linear combination $\tilde{S}_{\mu_n} = [\mathcal{P}^1_{\mu_n}+\mathcal{P}^0_{\mu_n}]S$. This object is then contracted with the gamma matrices and sequential propagator to construct $\mathcal{C}_{\mu_1\mu_2}$ for $n=2$. Two contractions are necessary for temporal $\mu_2$, and one for spatial. We then proceed to use these intermediate correlators to construct the full three-point function $C^{uu}_{\mu_1\mu_2}$ as defined in Eq.~\eqref{eq:Cefficient}. Starting with $\tilde{S}_{\mu_n}$, we then repeat the procedure with the next layer of covariant derivative, $\protect\overrightarrow{D}_{\mu_{n-1}}$ and so forth, until reaching the maximum $n$ desired, with the $\mu_i$ iterations of the intermediate $\mu_1\mu_2$, $\mu_1\mu_2\mu_3$, etc. contractions ordered depth first.}
    \label{fig:algo}
\end{figure}

In Fig.~\ref{fig:algo} we show a schematic representation of the procedure described above to optimize the computation of $\mathcal{C}_{\mu_1...\mu_n}$, including the implementation of the depth first ordering in the contractions.

\subsection{Measurement details}
\label{ssec:measurements}

Table~\ref{tab:ensembles} summarizes the setup of our measurements, including the ensembles analyzed, the source and sink parameters, and the range of flow times at which lattice correlation functions were evaluated.

A different number of independent gauge configurations was analyzed at each lattice spacing, chosen such that the relative uncertainties of the lowest nontrivial ratio $\braket{x^2}/\braket{x}$ remain approximately uniform across ensembles.  
For each configuration, three-point functions are computed using a single randomly chosen point source, with $\mathbbm{Z}_4$ stochastic noise applied to approximate volume averaging.  
Two sink-source time separations $\tau_s$ are employed on every ensemble to help control excited-state effects.  

To study the flow-time dependence of the correlation functions, measurements are performed in the range $t/t_0 \in [0,2.6]$, with spacings $\delta t/t_0 \simeq 0.05$--$0.1$ (see Table~\ref{tab:ensembles}).  
Statistical uncertainties and correlations are propagated using bootstrap resampling with $500$ samples per ensemble, applied simultaneously across all computed quantities: moments $n$, sink times $\tau_s$, operator insertion times $\tau_{\mcO}$, and flow times $t$.  
This setup ensures a consistent and controlled treatment of uncertainties throughout the analysis.

\section{Analysis strategy and results}
\label{sec:analysis_results}

Once the three-point functions of the twist-2 operators have been computed as described in Sec.~\ref{sec:lattice_setup}, several analysis steps are required to connect them to our target quantity: the ratios of PDF moments $\braket{x^{n-1}}/\braket{x}$ for $n=3,4,5,6$ in the $\overline{\text{MS}}$ scheme at the reference scale $\mu=2~\text{GeV}$, chosen to facilitate comparison with previous work. Specifically, we
\begin{itemize}
  \item extract the ratios of ground-state matrix elements $\braket{\pi(\mathbf{0})|\widehat{O}_{n,4\ldots4}(t)|\pi(\mathbf{0})}/\braket{\pi(\mathbf{0})|\widehat{O}_{m,4\ldots4}(t)|\pi(\mathbf{0})}$, with $n>m$, from the three-point functions at fixed flow time $t$; ratios of matrix elements at fixed $t$ admit a well-defined continuum limit,
  \item take the continuum limit of these ratios at each fixed $t$ using the four ensembles with lattice spacings $a\simeq\{0.12,\,0.094,\,0.077,\,0.064\}~\text{fm}$,
  \item multiply by the short-flow-time matching coefficients and fit any residual $t$ dependence to perform the $t\to 0$ extrapolation, thus obtaining the $\MSbar$ results for $\braket{x^{n-1}}/\braket{x}$ at $\mu=2~\text{GeV}$.
\end{itemize}
We address these steps in turn in the subsections below.

\subsection{Excited-state contamination}
\label{ssec:excited_states}

\begin{figure}[t]
    \centering
    \includegraphics[width=0.99\textwidth]{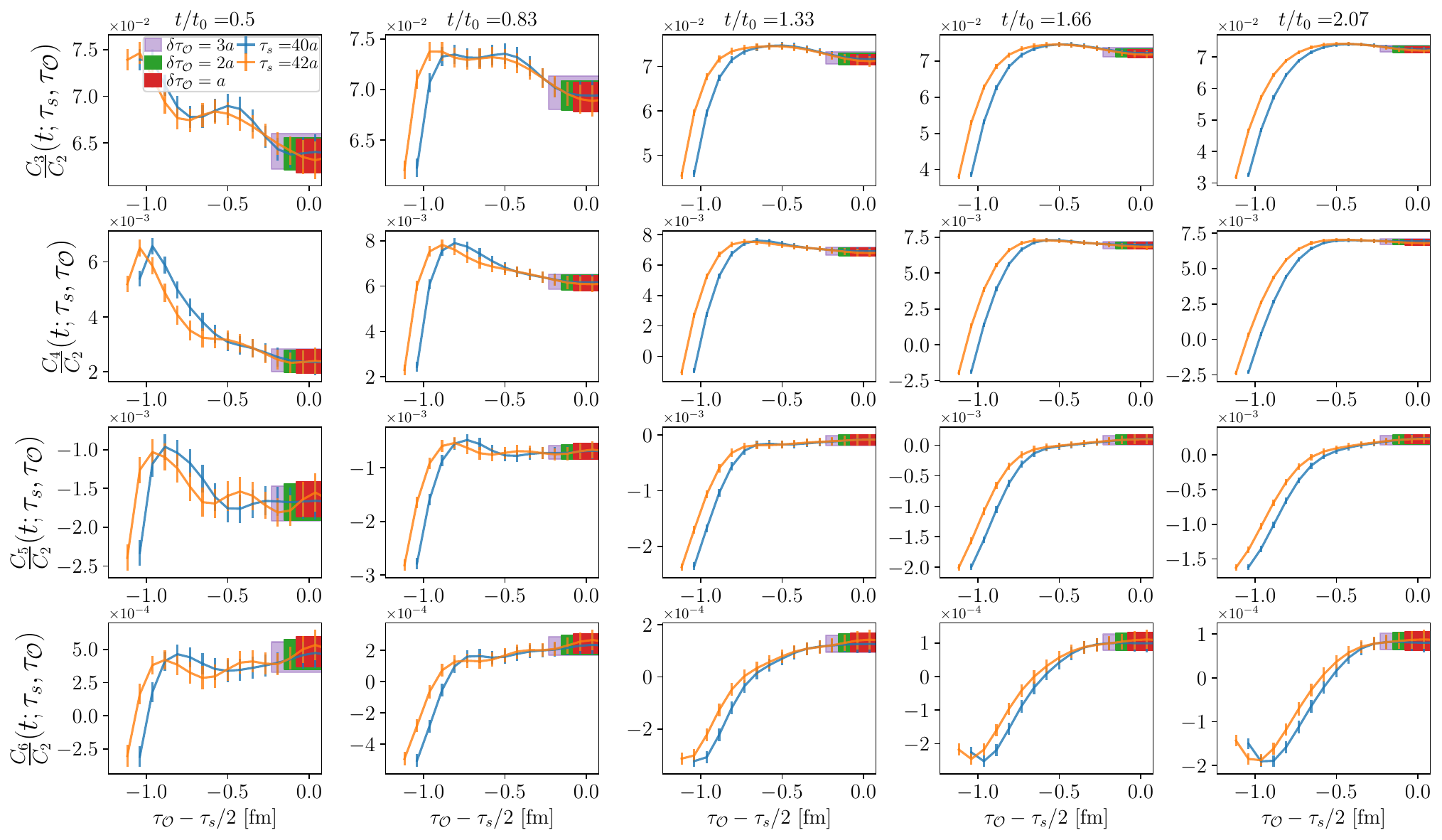}
    \caption{Examples of three-point function ratios $C_n/C_2$ for all $n$ and $\tau_s$ on ensemble \texttt{a077}, shown at five different flow times. For each case, we plot the corresponding effective estimator for the ground-state matrix element ratio, as described in \textit{Method 1}. We compare three averaging intervals, $\delta \tau_{\mcO}=a$ (red band), $2a$ (green band), and $3a$ (lilac band).}
    \label{fig:noplateaus_window}
\end{figure}

When extracting the ground-state matrix elements from the three-point functions of the twist-2 operators $\widehat{O}_{\mu_1...\mu_n}(t)$, it is important to assess possible excited-state contaminations (ESC).  
The flowed operator acquires a physical ``extension'' characterized by the flow-time radius $\sqrt{8t}$.  
In the region where this scale is much smaller than the separation from the pion creation and annihilation operators,  
$\sqrt{8t} \ll \tau_s - \tau_{\mcO}$ and $\sqrt{8t} \ll \tau_{\mcO}$, one can express the corresponding three-point function through the spectral decomposition
\be
C_n(\tau_s,\tau_{\mcO}; t) = 
\sum_{k,k'=0}^{\infty} 
\frac{e^{-E_{k'}\tau_s} e^{-(E_k-E_{k'})\tau_{\mcO}}}{4 E_k E_{k'}} 
Z_k^* Z_{k'} 
\braket{k'(\bzero)|\widehat{O}_n(t)|k(\bzero)} ,
\label{eq:spectral}
\ee
where the sum runs over all states $k$ in the spectrum, with energies $E_k$ and overlap factors $Z_k \equiv \braket{0|P^{ud}|k(\bzero)}$.  
Here, $C_n$ denotes the three-point function constructed from the \emph{symmetrized and traceless} twist-2 operator
$\widehat{O}_n(t) \equiv \widehat{O}_{\underbrace{\scriptstyle{4 \cdots 4}}_{n}}(t)$.

Since our goal is to determine ratios of ground-state matrix elements,
\be
\frac{\braket{\pi(\bzero)|\widehat{O}_{n}(t)|\pi(\bzero)}}
{\braket{\pi(\bzero)|\widehat{O}_{m}(t)|\pi(\bzero)}} ,
\label{eq:matrixratio}
\ee
for which the flowed fermion field renormalization cancels, the simplest approach is to form ratios of three-point functions for each $t$, $\tau_s$, and $\tau_{\mcO}$, which at sufficiently large Euclidean times $\tau_s-\tau_{\mcO}\gg 0$ and $\tau_{\mcO}\gg 0$ approach
\be
\lim_{\tau_{\mcO},\,\tau_s-\tau_{\mcO}\to\infty}
\frac{C_n(\tau_s,\tau_{\mcO};t)}{C_m(\tau_s,\tau_{\mcO};t)}
= 
\frac{\braket{\pi(\bzero)|\widehat{O}_{n}(t)|\pi(\bzero)}}
{\braket{\pi(\bzero)|\widehat{O}_{m}(t)|\pi(\bzero)}} 
+ \ldots ,
\label{eq:threepointratio}
\ee
where the ellipsis represents exponentially suppressed contributions from excited states.

A detailed analysis is therefore necessary to ensure that such contaminations are properly controlled.  
In addition to using the ratio method, we also consider direct multi-state fits to the three-point functions based on the truncated spectral decomposition in Eq.~\eqref{eq:spectral}, with $k,k'\leq k_{\text{max}}$.  

Motivated by this, we investigate three different procedures for extracting the desired matrix elements:
\begin{itemize}
    \item \textit{Method 1}: Use three-point function ratios as effective estimators of the matrix element ratios.
    \item \textit{Method 2}: Perform plateau fits to the three-point function ratios to extract the matrix element ratios.
    \item \textit{Method 3}: Use multi-state fits to extract matrix elements for each $t$, and then take ratios between $n$ and $m$ to cancel the flowed fermion field renormalization.
\end{itemize}

In all cases, taking $m>2$ in the denominator offers no advantage: the statistical uncertainty increases with $m$, while the control over the ratios does not improve.  
We therefore fix $m=2$ for the remainder of the analysis.  
Moreover, since the three-point functions at zero momentum are expected to be symmetric under $\tau_{\mcO} \rightarrow \tau_s - \tau_{\mcO}$ in the infinite-statistics limit, we enforce this symmetry explicitly by averaging them around the midpoint $\tau_{\mcO}=\tau_s/2$.

\subsubsection{Method 1: Ratios as effective estimators}

Eq.~\eqref{eq:threepointratio} implies that, in certain regions of parameter space, the ratio of three-point functions can be interpreted as an effective estimator for the ratio of ground-state matrix elements.  
The optimal choice of parameters that saturates the limit in Eq.~\eqref{eq:threepointratio} within our dataset is $(\tau_s,\tau_{\mcO}) = (\tau_{s,\mathrm{max}},\,\tau_{s,\mathrm{max}}/2)$, where $\tau_{s,\mathrm{max}}$ is the maximum available sink time for each ensemble.  

To account for statistical fluctuations, we average over all $(\tau_s,\tau_{\mcO})$ within a window around the midpoint,  
$\tau_s/2-\delta \tau_{\mcO} \leq \tau_{\mcO} \leq \tau_s/2+\delta \tau_{\mcO}$,  
and check the sensitivity of the results to the choice of $\delta \tau_{\mcO}$.  
The systematic uncertainty is set to half of the difference between the minimum and maximum central values of the ratios being averaged together.  

In Fig.~\ref{fig:noplateaus_window}, we show examples of the averaged estimators for three different choices of $\delta \tau_{\mcO}$ for one of the ensembles, plotted against the corresponding three-point function ratios.  
Equivalent behavior is observed across other ensembles and flow times.  
Given the weak dependence of the results on $\delta \tau_{\mcO}$, we fix $\delta \tau_{\mcO}=2a$ for the final analysis.

\subsubsection{Method 2: Plateau fits to ratios}

As expected from Eq.~\eqref{eq:threepointratio}, the ratio of three-point functions approaches a constant in the plateau region up to ESC.  
Accordingly, we fit the ratios to a constant by performing simultaneous fits to both values of $\tau_s$, applied to all contiguous $\tau_{\mcO}$ intervals satisfying  
$\tau_{\mcO,\text{min}} \leq \tau_{\mcO} \leq \tau_s - \tau_{\mcO,\text{min}}$,  
where $\tau_{\mcO,\text{min}}/a = 8,\,10,\,12,\,14$ for ensembles \texttt{a12}, \texttt{a094}, \texttt{a077}, and \texttt{a064}, respectively.  
Here $\tau_{\mcO,\text{min}}$ corresponds to the minimal distance of the flowed operator from both the source and the sink, and it is chosen to be approximately constant in physical units, about $0.9$-$0.95~\text{fm}$, across the four lattice spacings.  
This minimum separation is selected to exclude regions where both excited-state and finite-flow-radius effects are expected to be significant.

\begin{figure}[t]
    \centering
    \includegraphics[width=0.99\textwidth]{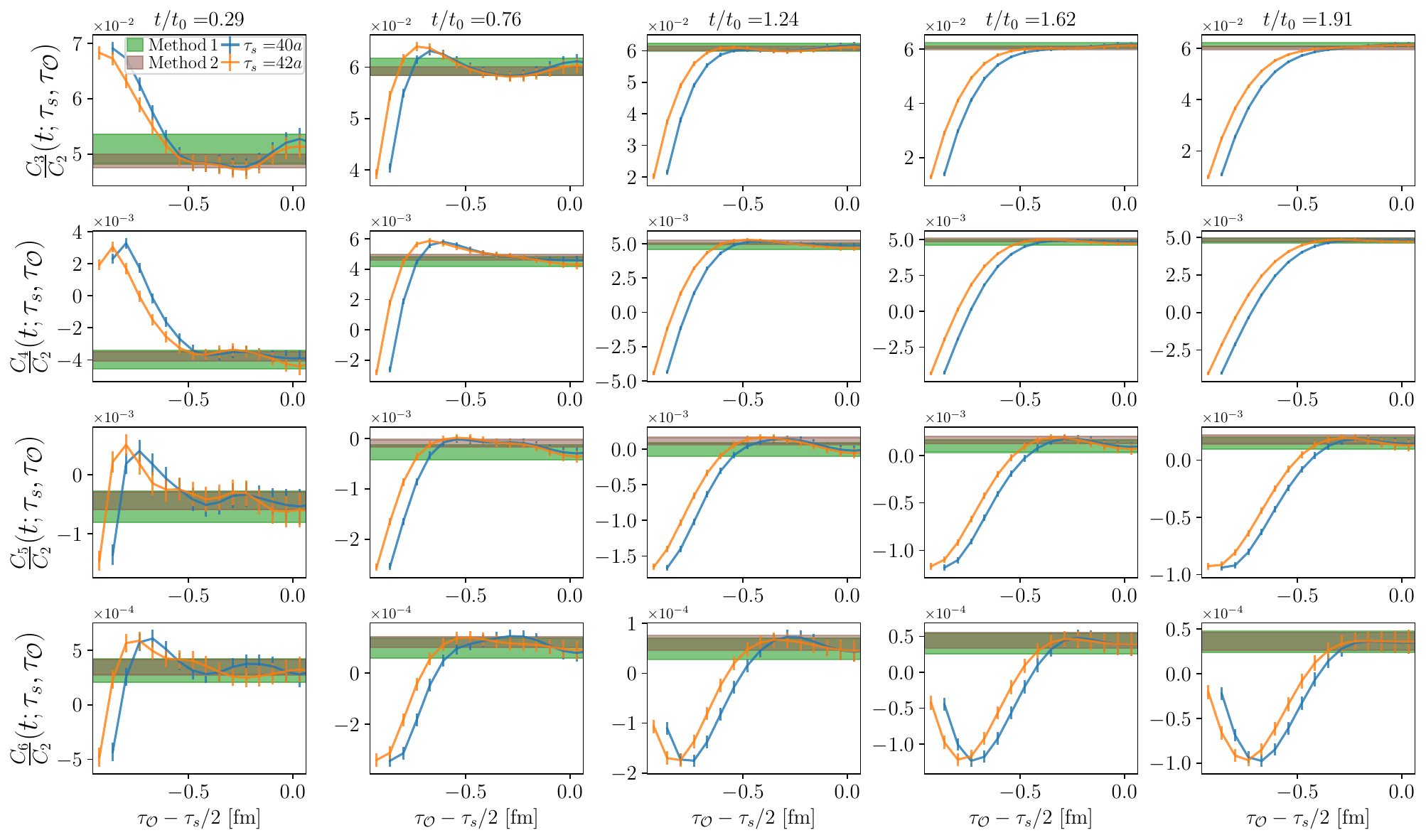}
    \caption{Examples of three-point function ratios for the {\tt a064} ensemble, shown as in Fig.~2.
    The green band corresponds to the effective estimator (\textit{Method 1}) with $\delta\tau_{\mcO}=2a$.
    The brown band represents the AIC-weighted average of plateau fits over several regions, as described in \textit{Method 2}.
}
    \label{fig:method12compare}
\end{figure}

All plateau fits that include at least eight data points are combined through a weighted average based on their Akaike Information Criterion (AIC) weights~\cite{Akaike:1998zah,Jay:2020jkz}.  
The resulting matrix element ratios are consistent with, but statistically more precise than, the effective estimators obtained with Method~1, as shown in Fig.~\ref{fig:method12compare}.

\begin{figure}[h]
    \centering
    \includegraphics[width=0.9\textwidth]{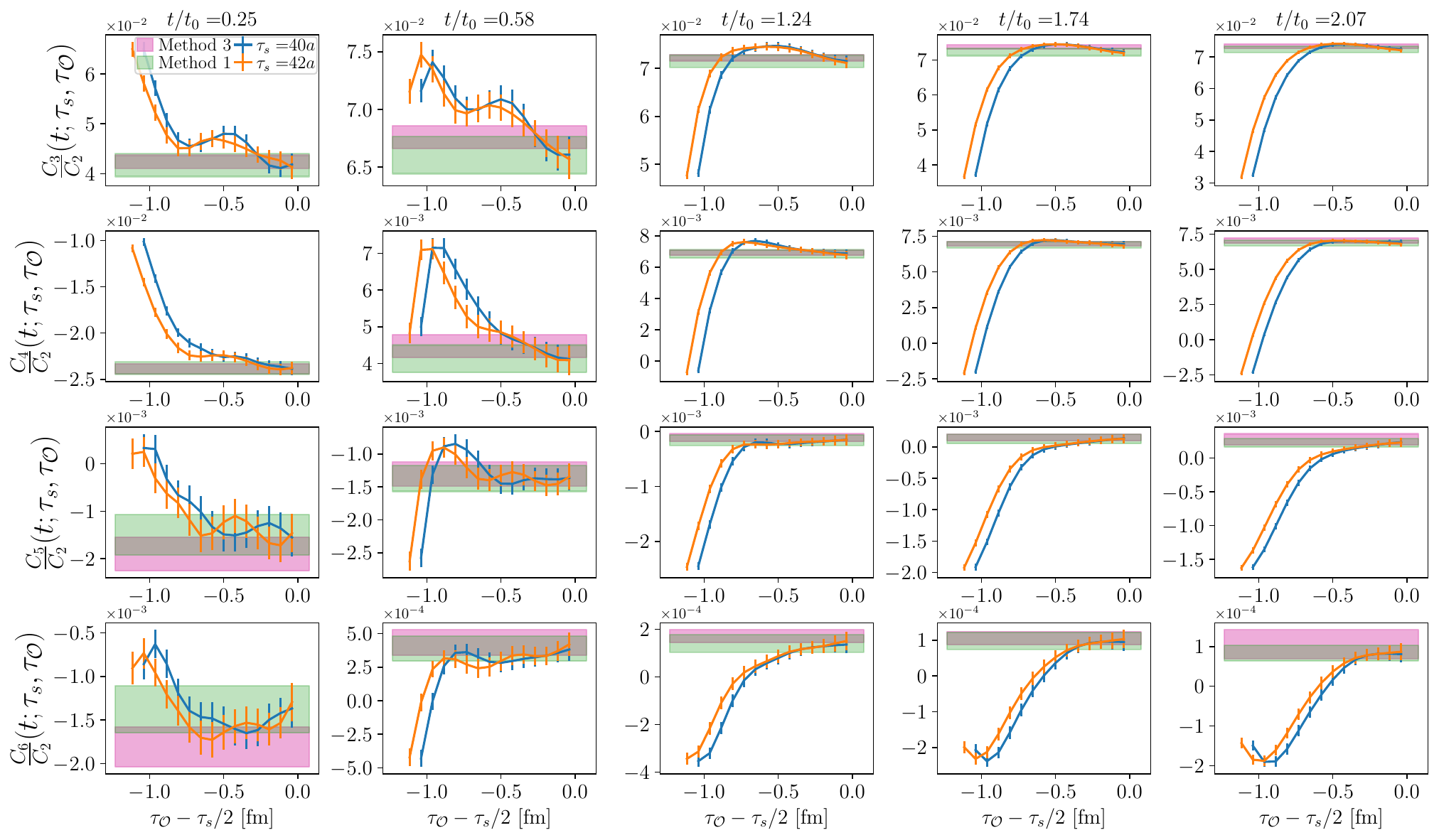}
    \caption{Examples of excited-state fit results (\textit{Method 3}) for ensemble {\tt a077} and five different flow times.
    For each flow time, the fit includes three states and is performed simultaneously to the corresponding two-point function together with all three-point functions $C_n(t; \tau_s,\tau_{\mcO})$ for all $n$ and both sink times, $\tau_s$.
    From these fits we obtain the ground-state matrix elements for all $n$, and divide those with $n>2$ by the $n=2$ matrix element. The results are shown here as the pink band, plotted against the corresponding three-point function ratios ($\tau_s = 40a$ with blue points and $\tau_s = 42a$ with orange points).
    For comparison, the effective estimator (\textit{Method 1}) with $\delta\tau_{\mcO}=2a$ is also included (green band).} 
    \label{fig:method13compare}
\end{figure}

\subsubsection{Method 3: Excited-state fits}

Finally, we attempt to extract the matrix elements directly from excited-state fits to the three-point functions, rather than forming ratios first.  
This approach provides a more detailed handle on possible excited-state contamination and serves as a cross-check of the assumptions underlying the previous methods.

Although the spectra $E_k$ and overlaps $Z_k$ defined in Eq.~\eqref{eq:spectral} are common to all flow times, the limited statistics of our ensembles (400-838 configurations) do not allow for a stable determination of the full covariance matrix required for a global simultaneous fit.  
We therefore perform separate fits at each $t$, accounting for correlations between flow times via bootstrap resampling.  
The three-point functions at both $\tau_s$ are fit simultaneously with the unflowed pion two-point correlation function in order to better constrain the energy levels and overlap factors.  
We consider $k_{\text{max}}=1$ and $2$ (corresponding to two- and three-state fits), varying the range of $\tau_{\mcO}$ included in the fit, while the minimum and maximum time slices of the two-point function are varied independently.  

The excited-state fits are typically less stable and more computationally demanding than the plateau analyses.  
In principle, Bayesian priors could help stabilize the fits; however, given the present data, which show minimal apparent excited-state contamination, we do not expect a significant improvement.  
In this proof-of-principle study, we therefore restrict the analysis to a few representative flow times $t$ for each ensemble.  
Within the current statistical precision, these fits yield ground-state matrix element ratios consistent with those obtained using Method~1, as illustrated for selected examples in Fig.~\ref{fig:method13compare}.  
In future work, with higher-precision data and additional source-sink separations, or in studies of nucleon moments where the ground state is more difficult to isolate, we plan to explore extended analysis strategies.

\vspace{0.5em}
All three methods (effective estimators, plateau fits, and excited-state fits) produce consistent results, demonstrating that within our current statistical precision, excited-state contaminations do not affect the final matrix element ratios.  
We therefore adopt the conservative choice of using Method~1 for the final analysis, as it provides results insensitive to the small statistical fluctuations observed in some plateaus, at the cost of slightly larger statistical uncertainties.

\subsection{Continuum limit}
\label{ssec:continuum_limit}

\begin{figure*}[t]
\centering
    \includegraphics[width=0.99\linewidth]{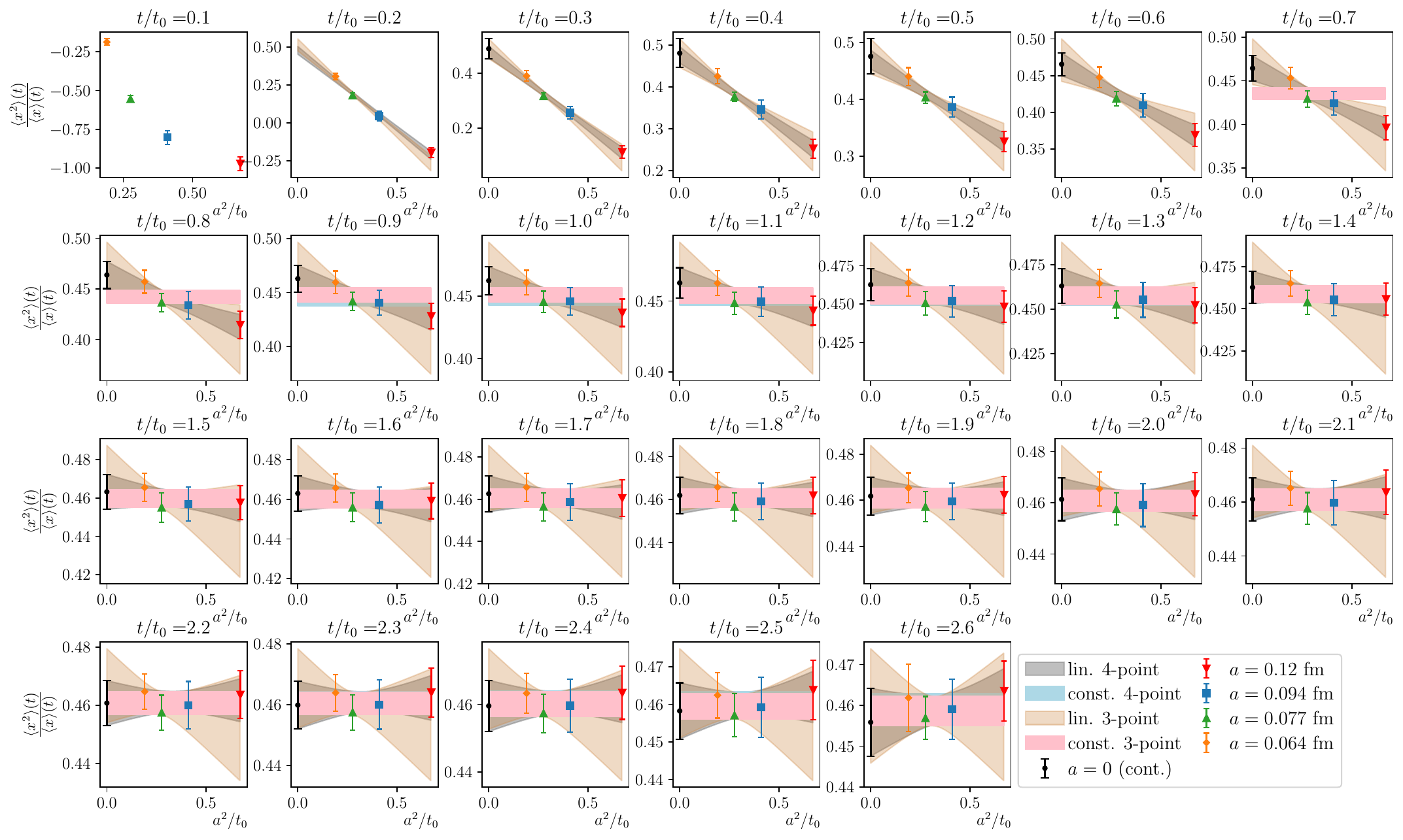}
    \centering
    \includegraphics[width=0.99\linewidth]{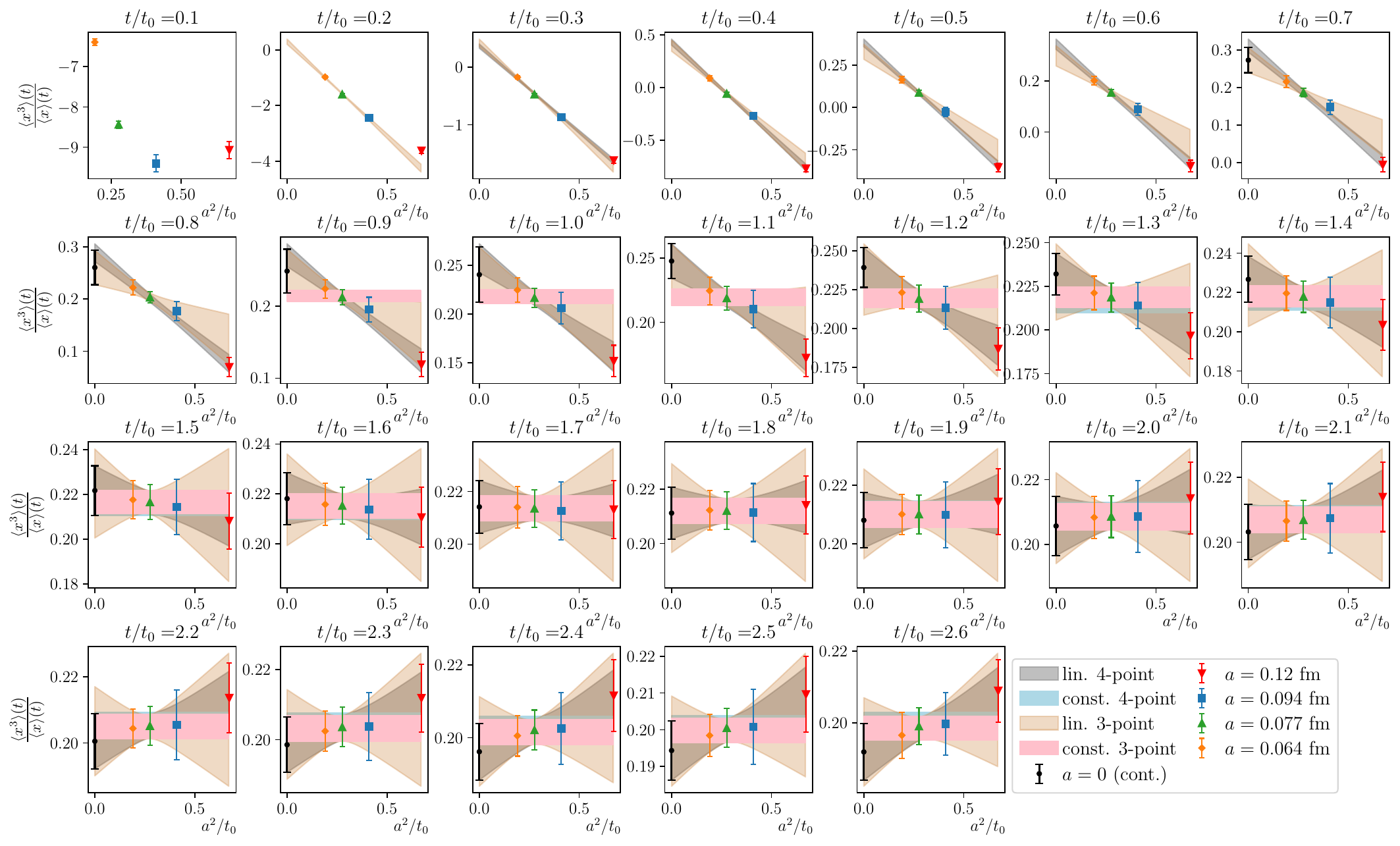}
    \caption{
    Investigation of the continuum extrapolation for all flow times of $n=3$ (top panel) and $n=4$ (bottom panel).  
    Four different fits are considered, as described in the text, and each is shown only when its reduced $\chi^2_\nu < 1.5$.  
    The shaded bands indicate the corresponding fit type, while the black data points mark the continuum-limit results obtained with the fit function adopted for the final analysis.  
    These are shown for those values of $t/t_0$ where the continuum extrapolation is considered robust, according to the criteria discussed in the text.}
    \label{fig:x2x3textrap}
\end{figure*}
\begin{figure*}[t]
    \centering
    \includegraphics[width=0.99\linewidth]{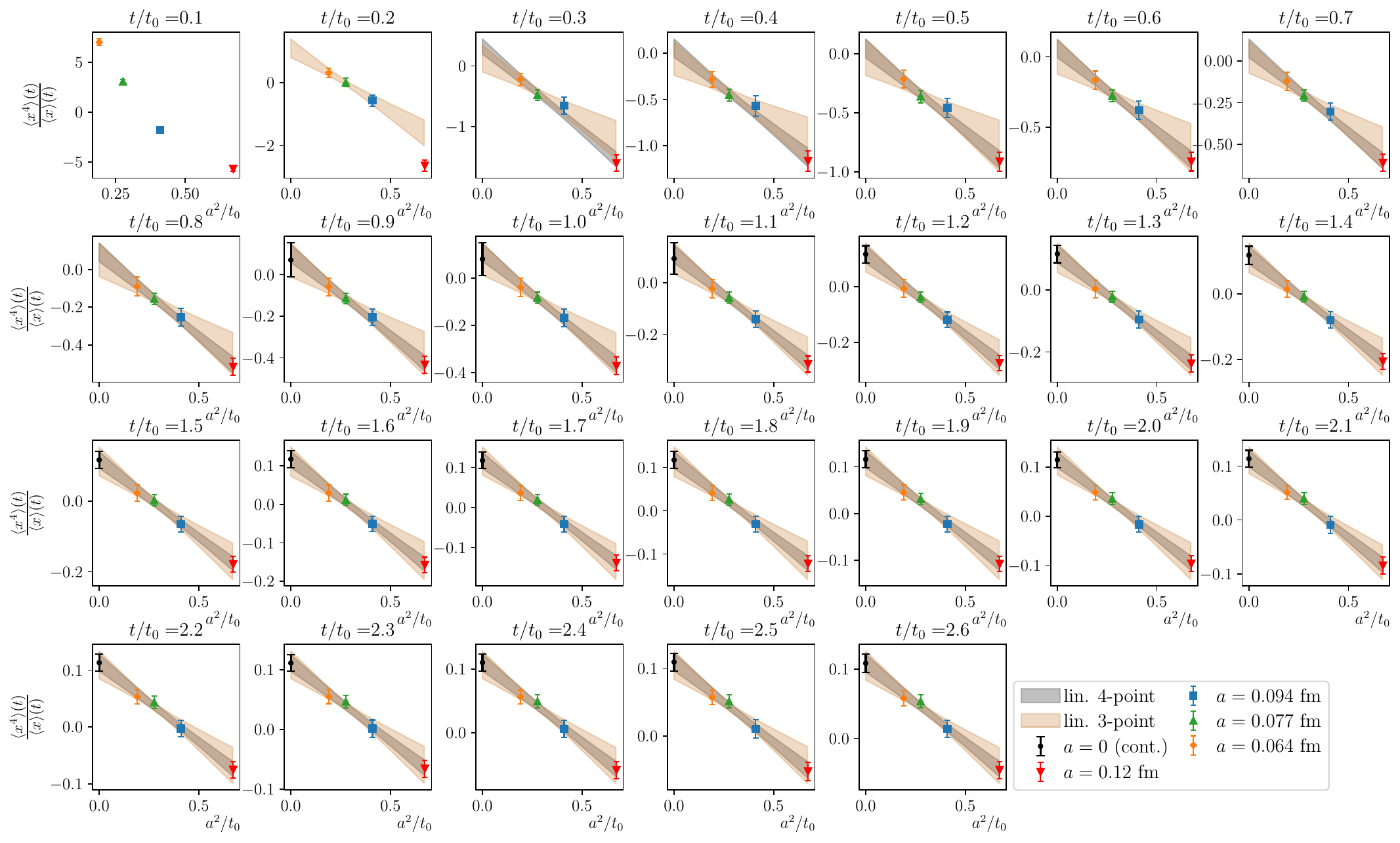}
     \centering
    \includegraphics[width=0.99\linewidth]{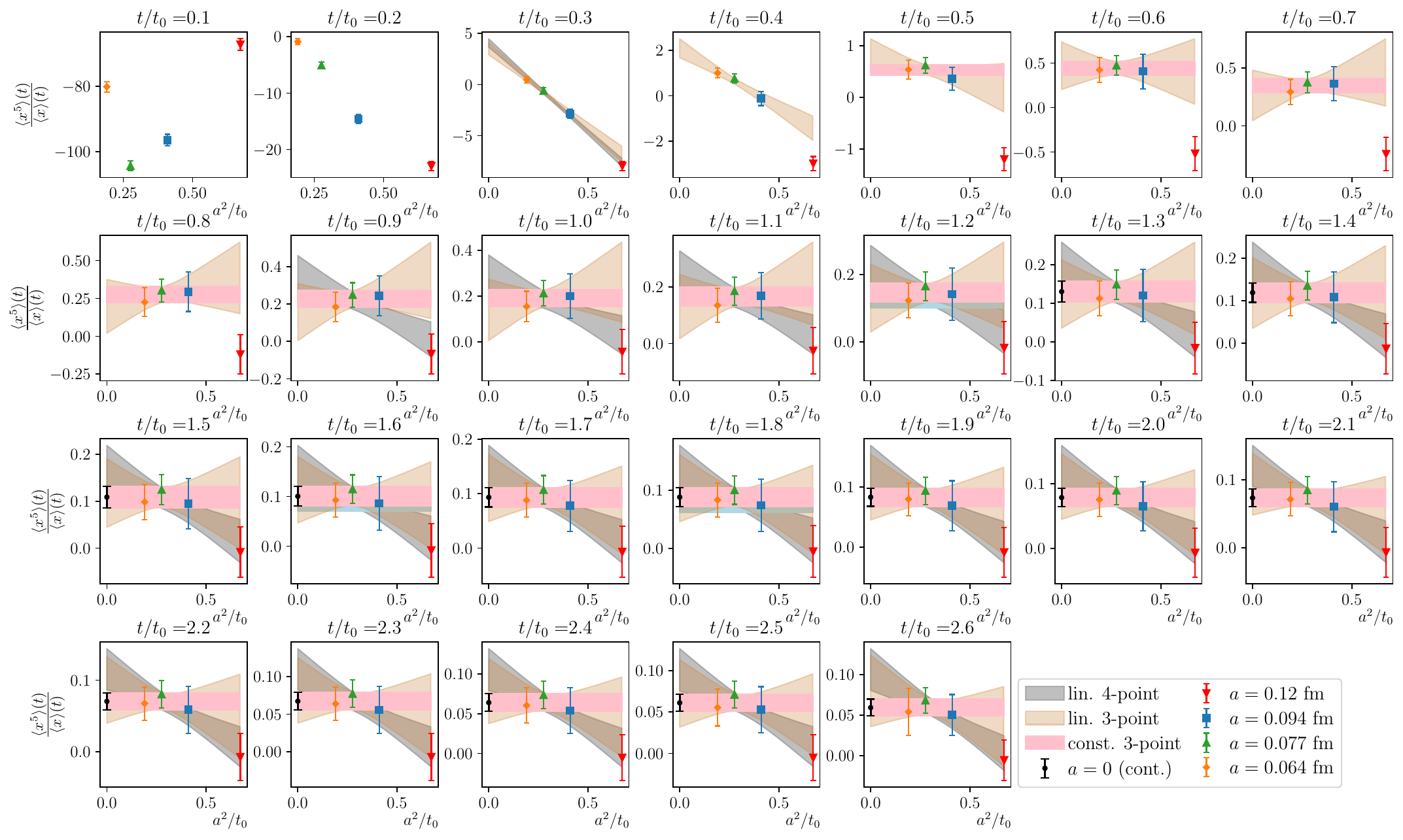}
    \caption{Same as Fig.~\ref{fig:x2x3textrap}, but for $n=5$ (top panel) and $n=6$ (bottom panel).}
    \label{fig:x4x5textrap}
\end{figure*}

Having determined the matrix element ratios for all ensembles and flow times, we can proceed with the continuum-limit extrapolation.  
We work directly with the ratios rescaled by the appropriate powers of the pion mass and factors of $(-1)$ corresponding to the flowed moment ratios, i.e.,
\be 
r_n
= \frac{(-1)^{n}}{(am_{\pi})^{\,n-2}}
  \frac{\braket{\pi({\bf 0})|\widehat{\mcO}{n}(t)|\pi({\bf 0})}}
       {\braket{\pi({\bf 0})|\widehat{\mcO}{44}(t)|\pi({\bf 0})}} .
\label{eq:momentratioflow}
\ee
The continuum limit must be taken at fixed values of $t/t_0$; therefore, we first interpolate the flowed moment ratios of each ensemble with respect to $t/t_0$ using cubic splines.  
This yields values for all ensembles on a common grid of $t/t_0 \in [0.1,\,2.7]$ with spacing $\delta(t/t_0)=0.1$, chosen close to the original measurement spacing.
The lattice action used in this work, the expClover action, employs a nonperturbatively determined $\csw$ coefficient~\cite{Francis:2019muy,Francis:2022hyr}, ensuring O($a$) improvement of spectral quantities.  
Possible residual O($a$) effects could therefore only arise from the local flowed operator $\mcO(t)$.  
As shown in Refs.~\cite{Luscher:2013cpa,Shindler:2013bia}, lattice and chiral symmetries restrict the appearance of such effects to three cases.  
The first is an additional dimension-5 operator in the Symanzik effective theory, which contributes only when two flowed fermion fields are contracted; since in our setup flowed fields are always contracted with unflowed ones, this term is absent.  
The second source consists of universal O($am$) effects that modify any fermionic operator through the flowed fermion field renormalization.  
These affect all operators with the same fermion content in the same way and thus cancel exactly in the ratios of flowed moments.  
Finally, short-distance cutoff effects can arise when the flowed and unflowed operators in the correlation functions approach each other in physical distance.  
These effects can be represented by correlation functions in which one of the quark propagators attached to the flowed operator is replaced by the kernel $K(t;y,0)$ obtained by solving the discretized GF equation with the proper source as an initial condition.  
We have explicitly verified, for representative gauge configurations, that the ratio of these O($a$) correlation functions to the connected three-point functions used to determine the flowed moments,  
is negligible in the plateau region, with deviations appearing only at very short distances.  
This demonstrates that the ratios used in this analysis are effectively O($a$)-improved.

Finally, by dimensional arguments we expect that O($a^2$) effects may still exhibit a dependence on the flow time, scaling as $a^2/t$.  
We therefore allow the O($a^2$) coefficients to depend on $t$ in our continuum extrapolation fits.  
The leading discretization effects are then parametrized as
\be
r_n(a^2/t_0,t/t_0) = \frac{\braket{x^{n-1}}}{\braket{x}}\left(t/t_0\right) + \delta_{n,2}(t/t_0)\,\frac{a^2}{t_0} + \cdots \, .
\label{eq:cont_a2_form}
\ee
Our strategy is to extrapolate each moment ratio to the continuum limit independently at every flow time, using for each $t$ the set of lattice spacings that provide a stable and well-described fit according to Eq.~\eqref{eq:cont_a2_form}.  
In this approach, two flow-time cuts must be defined for each moment $n$:  
(1) the minimum $\left(t/t_0\right)_{n,\text{min}}$, below which the data are not consistent with the assumed discretization form and are therefore discarded; and  
(2) the threshold $\left(t/t_0\right)_{n,3\leftarrow 4}$ below which the coarsest lattice spacing is excluded from the fit.  
In addition, we must decide whether to include the linear $a^2$ term in the extrapolation and for which values of $n$ and $t$ this is justified.

To determine these analysis choices and to assess their impact on the final results, we perform four types of continuum extrapolations for each $n$ and $t$:  
(1) a linear fit in $a^2$ using all four ensembles (``4-point linear''),  
(2) a linear fit using the finest three ensembles (``3-point linear''),  
(3) a constant fit using all four ensembles (``4-point constant''), and  
(4) a constant fit using the finest three ensembles (``3-point constant'').  
All fits with a reduced $\chi^2_\nu<1.5$ are displayed in Figs.~\ref{fig:x2x3textrap} and~\ref{fig:x4x5textrap} as colored bands.
For all moments $n$, we observe the expected pattern that cutoff effects increase as $t/t_0$ decreases.  
At very small flow times, the data show no clear scaling consistent with the Symanzik effective description.  
For slightly larger $t/t_0$, the coarsest lattice spacing departs from the scaling trend defined by the finer lattices, while at still larger $t/t_0$ the cutoff effects become mild and are well described by Eq.~\eqref{eq:cont_a2_form}.

Based on these observations, for $n=3$, we employ a 3-point linear fit once the {\tt a094} ensemble deviates by less than $40$-$50\%$ (corresponding to $0.3 \le t/t_0 \le 0.5$), and switch to a 4-point fit once the {\tt a12} ensemble falls below the same threshold ($t/t_0 > 0.5$).  
The same criterion is applied for $n=4$, resulting in a 3-point fit for $0.7 \le t/t_0 \le 1.0$ and a 4-point fit above $t/t_0 = 1.0$.  
For $n=5$, the continuum limit exhibits a noticeable $a^2$ slope but remains essentially independent of $t/t_0$.  
Here we start with a 3-point linear fit at $t/t_0 = 0.9$ (when {\tt a094} becomes usable) and switch to a 4-point fit at $t/t_0 = 1.1$ (when {\tt a12} enters).  
The specific choices are data-driven, and the consistently good $\chi^2_\nu$ values across all continuum extrapolations confirm the robustness of this procedure.  
Finally, for $n=6$, we exclude the {\tt a12} ensemble entirely because of strong cutoff effects and apply a 3-point constant fit starting where the slope between consecutive points drops from roughly $20\%$ to $15\%$ (around $t/t_0 = 1.3$).  

To help visualize the outcome of this analysis, Figs.~\ref{fig:x2x3textrap}-\ref{fig:x4x5textrap} show the resulting continuum limits for those $t/t_0$ values where full control of the extrapolation has been established.  
The shaded error bands correspond to the specific continuum extrapolation adopted, so the fit type used for the final results at each $n$ and $t$ can be identified directly from the band to which each point belongs.  
All fits yield excellent $\chi^2_\nu$ values, and the final ratios were verified to be stable under reasonable variations of these cuts.

\subsection{Extrapolation to $t \to 0$}
\label{ssec:t0_extrapolation}

\begin{figure*}[t]
    \centering
    \includegraphics[width=0.9\linewidth]{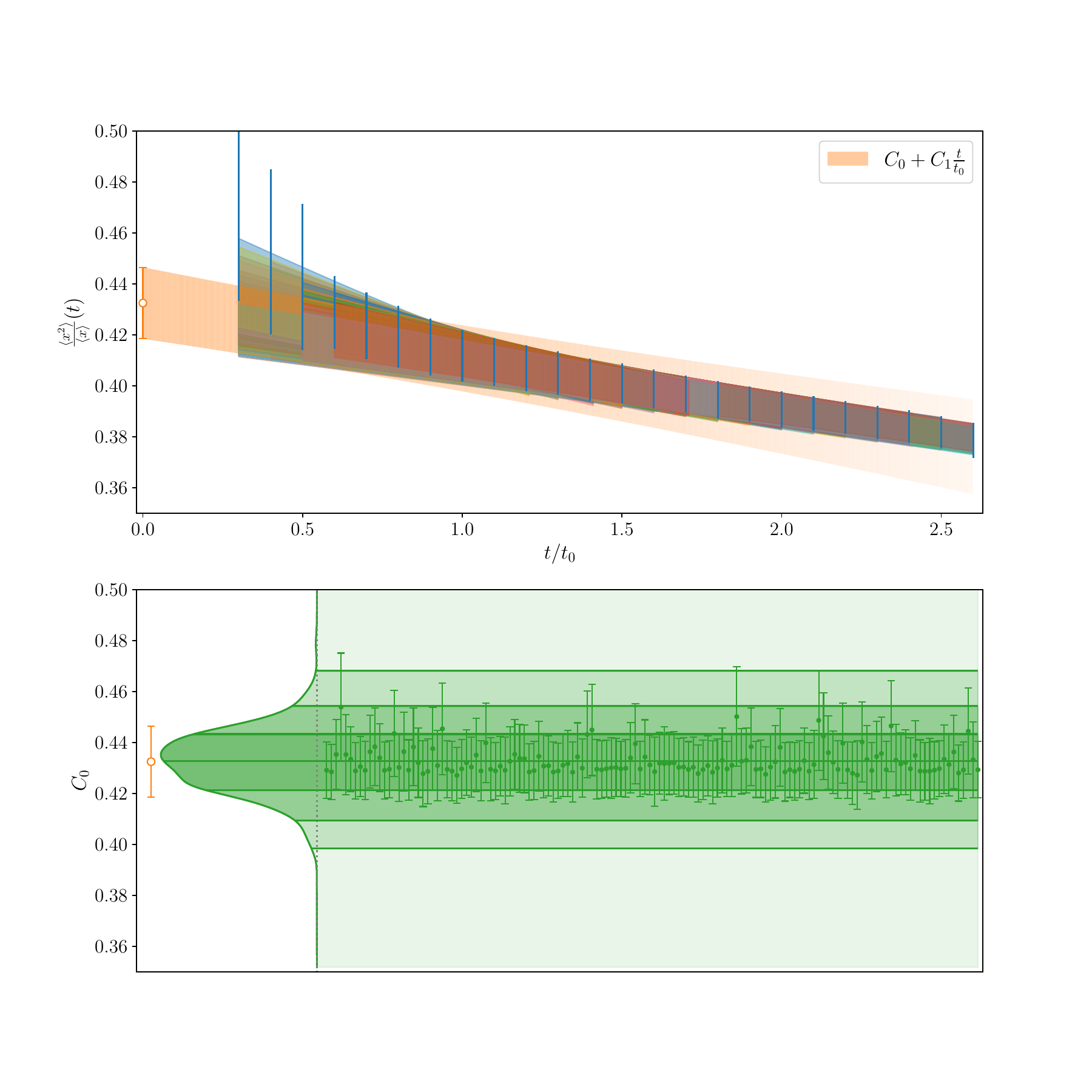}
\caption{
All individual flow-time fits contributing to the final result for $\braket{x^{2}}/\braket{x}$. 
\textbf{Top panel:} $t/t_0$ dependence of the moment ratio in the continuum limit and after matching (see Eq.~\eqref{eq:x_MS}), with individual linear fits $C_0 + C_1\, t/t_0$ shown in different colors. 
The fits shown are those accepted with $p$-values larger than $0.1$ and a minimum fit range $\Delta = 0.8$ within $[(t/t_0)_{\text{min}},\,2.6]$. 
The final result (orange band) is obtained by averaging over all accepted fits and adding in quadrature the systematic uncertainty, as described in the bottom panel and in the main text. 
\textbf{Bottom panel:} Distribution of the fit results for $\braket{x^{2}}/\braket{x}$ in the $\MSbar$ scheme at $\mu=2~\text{GeV}$, combining all bootstrap samples and fit-range variations, together with individual accepted fit results randomly ordered. 
The shaded areas indicate the $2.5$, $16$, $50$, $84$, and $97.5$ percentiles of the samples, while the orange error bar marks the symmetric $\pm1\sigma$ interval that includes the systematic uncertainty. 
The total uncertainty is obtained by adding in quadrature to the diagonal of the covariance matrix the variance of the central values of all fits with respect to their mean.
}
    \label{fig:x2x1tfits}
\end{figure*}

\begin{figure*}
    \centering
    \includegraphics[width=0.9\linewidth]{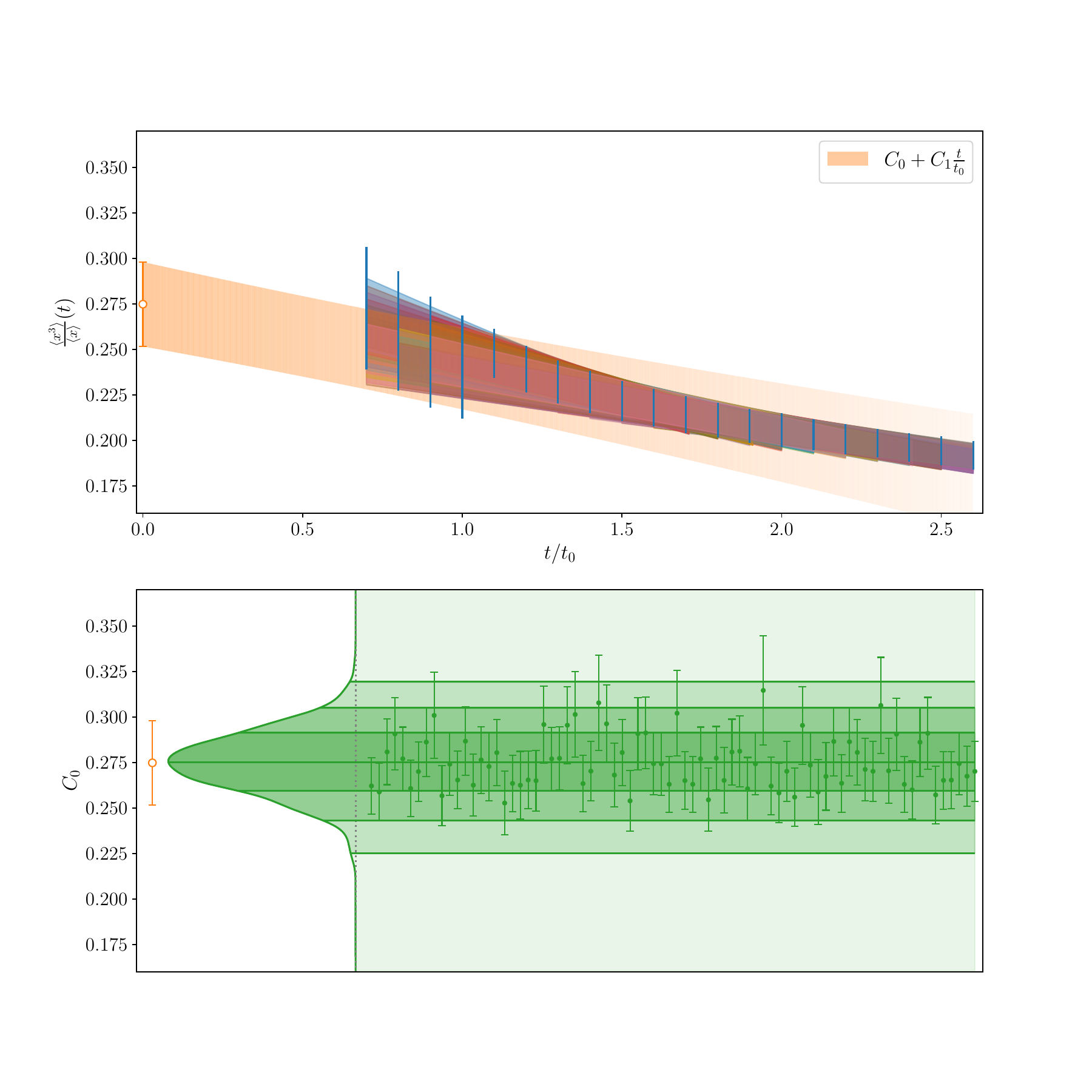}
    \caption{Same as Fig.~\ref{fig:x2x1tfits} for $\braket{x^3}/\braket{x}$. }
    \label{fig:x3x1tfits}
\end{figure*}
\begin{figure*}
    \centering
    \includegraphics[width=0.9\linewidth]{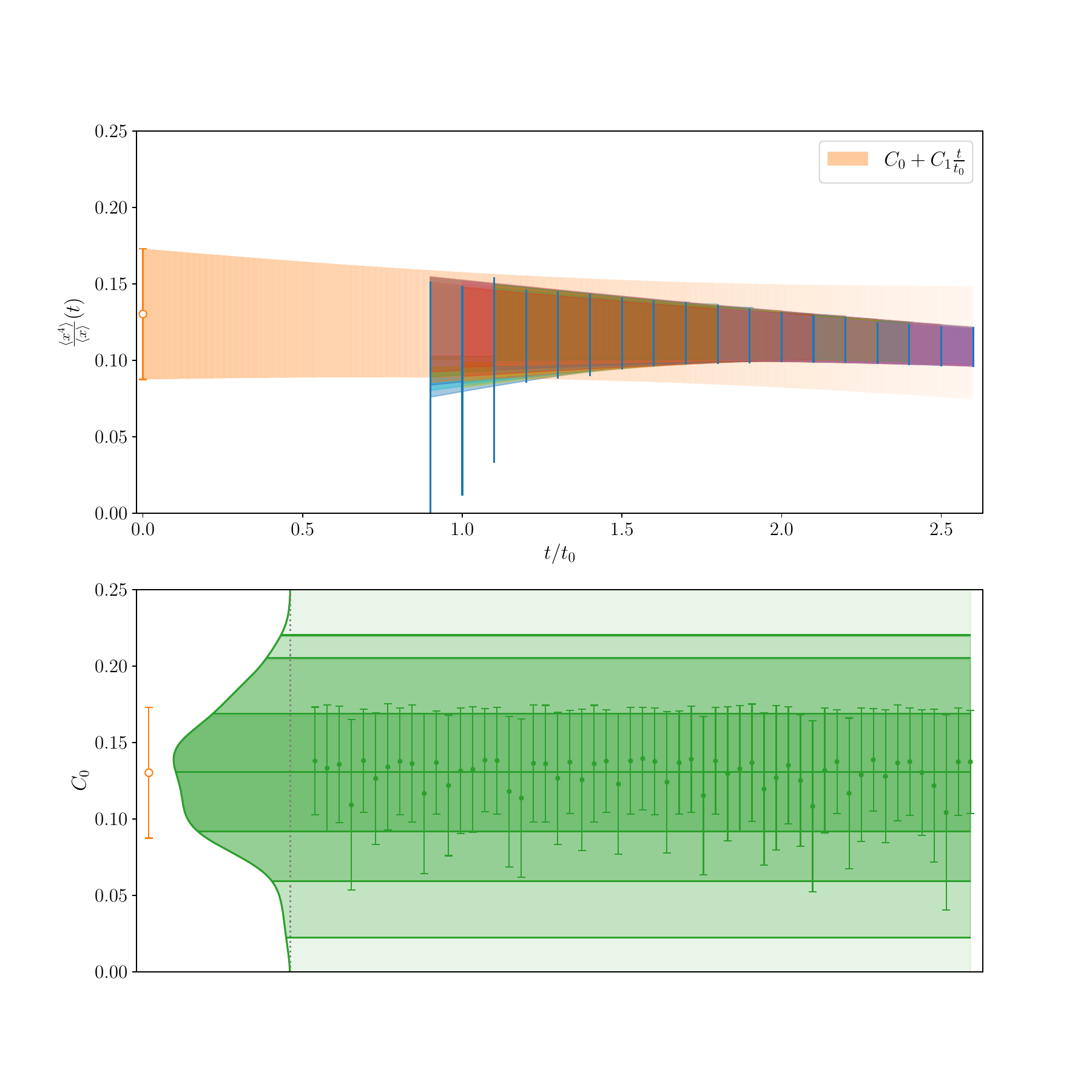}
    \caption{Same as Fig.~\ref{fig:x2x1tfits} for $\braket{x^4}/\braket{x}$.}
    \label{fig:x4x1tfits}
\end{figure*}
\begin{figure*}
    \centering
    \includegraphics[width=0.9\linewidth]{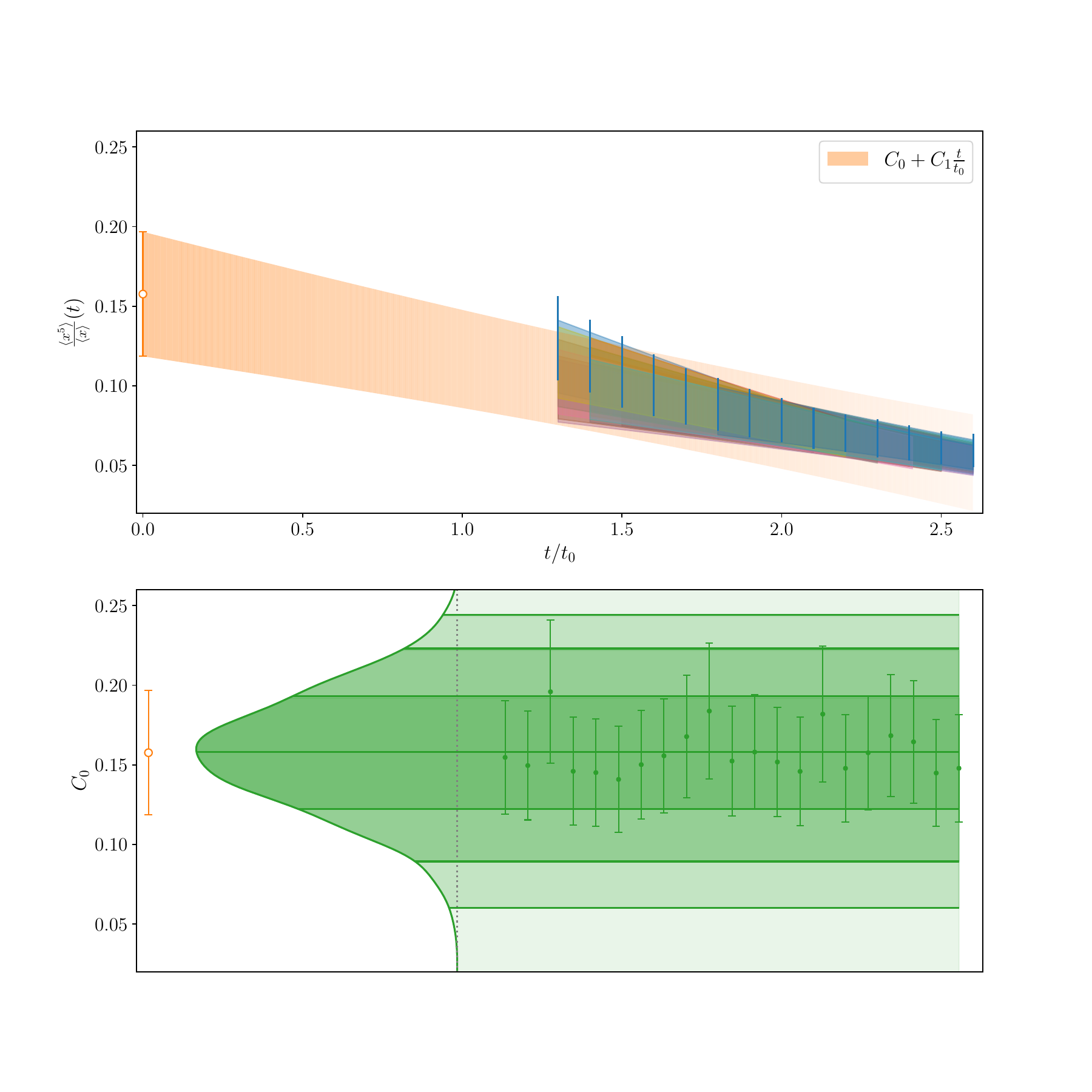}
    \caption{Same as Fig.~\ref{fig:x2x1tfits} for $\braket{x^5}/\braket{x}$. }
    \label{fig:x5x1tfits}
\end{figure*}

We now determine the ratios of moments in the $\overline{\text{MS}}$ scheme at the reference scale $\mu = 2~\text{GeV}$ by performing the extrapolation $t \to 0$.  
Our starting point is the continuum-extrapolated (i.e., $a \to 0$) flowed moment ratios obtained from the procedure described in Sec.~\ref{ssec:continuum_limit}.  
These continuum values are multiplied by the small-flow-time matching coefficients, $\zeta_n(t,\mu)$, which connect flowed renormalized quantities to their counterparts in the $\overline{\text{MS}}$ scheme, as described in Sec.~\ref{sec:new_method} (cfr. Eqs.~\eqref{eq:x_MS}).
The coefficients $\zeta_n(t,\mu)$ were computed up to NLO in Ref.~\cite{Shindler:2023xpd} and extended to next-to-next-to-leading order (NNLO) for the results presented here and in the accompanying Letter~\cite{Francis:2025rya}.  
A detailed account of the NNLO extension will be given in a forthcoming publication.  
The numerical impact of the matching coefficients, as well as the difference between NLO and NNLO, is discussed in Ref.~\cite{Francis:2025rya} and is therefore not repeated here.  
All subsequent analyses employ the NNLO-matched moment ratios.

After removal of discretization effects (Sec.~\ref{ssec:continuum_limit}) and application of the SFTX matching, the ratios can still exhibit a residual dependence on the flow time $t$.  
Such dependence may arise from higher-dimensional operators contributing at O($t$) or higher, and from logarithmic corrections associated with the truncation of the perturbative expansion of the matching coefficients.  
Within our current statistical precision, we find no evidence for logarithmic corrections and therefore adopt a simple linear ansatz in $t/t_0$ for the extrapolation to $t \to 0$.

As discussed in Sec.~\ref{ssec:continuum_limit}, the $a \to 0$ extrapolation imposes a lower bound on the flow time, $(t/t_0)_{\text{min}}$, for each moment $n$, determined by the region where the continuum limit is robust.  
The corresponding values are $\left(t/t_0\right)_{\text{min}} = [0.3,\,0.7,\,0.9,\,1.3]$ for $n = [3,\,4,\,5,\,6]$, respectively.  
For each $n$, we fit the data as a function of flow time using the linear ansatz 
$C_0 + C_1 \cdot t/t_0$
where $C_0$ and $C_1$ are fit parameters.  
Fits are performed over all flow-time intervals of width $\Delta \geq 0.8$ within the range $[(t/t_0)_{\text{min}},\,2.6]$, and only fits with $p$-values greater than $0.1$ are retained.

The final central value is obtained as a flat average of all accepted fits, while the systematic uncertainty is estimated as the mean squared deviation of the individual fit central values from this average.  
Figures~\ref{fig:x2x1tfits}-\ref{fig:x5x1tfits} illustrate this procedure, showing all individual linear fits over different flow-time intervals together with the resulting average and its total uncertainty.

Performing a weighted (AIC-based) average instead of a flat one leads to consistent but somewhat more precise results.  
We nevertheless adopt the flat average as a conservative choice: the AIC weights tend to favor fits with more degrees of freedom, which correspond to those dominated by data at larger $t$, where the SFTX is expected to gradually lose validity.  
We have verified that varying $\left(t/t_0\right)_{\text{min}}$ and $\left(t/t_0\right)_{\text{max}}$ within reasonable limits does not affect the results within uncertainties.  
We therefore conclude that our extrapolated values are stable up to $n=6$, the highest moment analyzed in this work.  
If the observed trend of extending cutoff effects with increasing $n$ persists, it may become necessary for higher moments ($n > 6$) to employ more elaborate combined $(a,t)$ extrapolations, e.g., including terms of order O($a^2/t$).  
We leave such investigations for future work.

The final results from the analysis just described for the moment ratios are summarized in Table~\ref{tab:results} and presented in the accompanying Letter~\cite{Francis:2025rya}.

\begin{table}[h]
\begin{center}
\begin{tabular}{c|cccc}
\hline\hline
  $n$ & 3  & 4 & 5 & 6 \\
\hline
 $~\braket{x^{n-1}}/\braket{x}~$ 
   & ~0.433(14)~ & ~0.275(23)~ & ~0.130(43)~ & ~0.158(39)~ \\
\hline\hline
\end{tabular}
\caption{\label{tab:results} 
Final results for $\braket{x^{n-1}}/\braket{x}$ in the $\overline{\MS}$ scheme at $\mu=2~\text{GeV}$.}
\end{center}
\end{table}

\section{Comparison with phenomenology and other lattice results}
\label{sec:comparison}

\begin{figure}[t]
    \centering
    \includegraphics[width=0.9\linewidth]{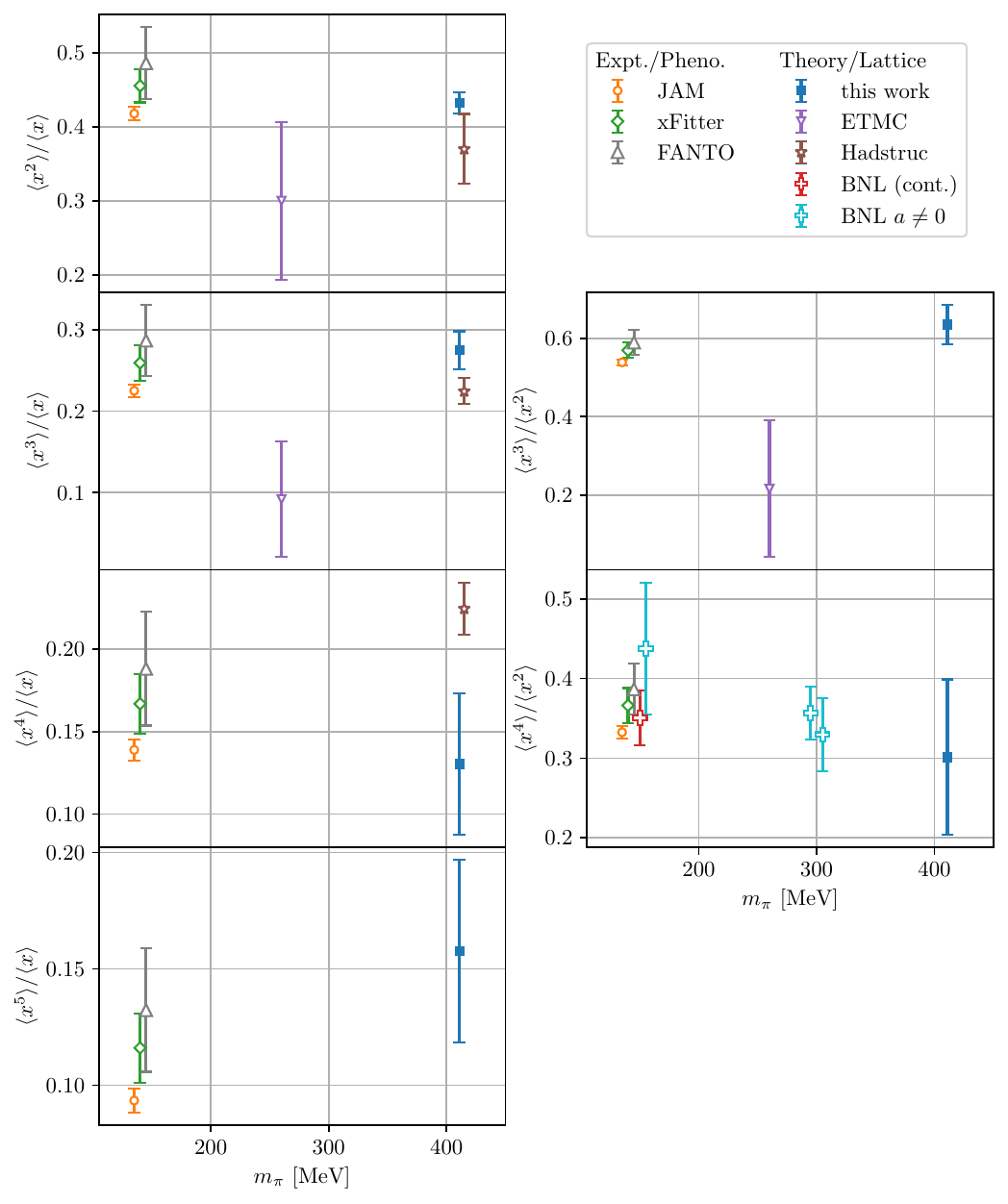}
    \caption{Comparison of our continuum-extrapolated results for the pion PDF moment ratios $\braket{x^{n-1}}/\braket{x}$ with recent phenomenological extractions~\cite{Barry:2021osv,Novikov:2020snp,Kotz:2025lio} and lattice QCD determinations~\cite{Alexandrou:2020gxs,Alexandrou:2021mmi,Gao:2022iex,Joo:2019bzr}. 
    All results are shown as a function of the pion mass $m_\pi$ and are renormalized in the $\MSbar$ scheme at the reference scale $\mu = 2~\text{GeV}$.
}
    \label{fig:mpicomparison}
\end{figure}

There is a long history of phenomenological extractions of the pion PDFs from experimental data~\cite{Owens:1984zj,AURENCHE1989517,Sutton:1991ay,Gluck:1991ey,Gluck:1999xe,Wijesooriya:2005ir,Aicher:2010cb,Barry:2018ort,Barry:2021osv,Novikov:2020snp,Kotz:2023pbu,Kotz:2025lio}.  
Although the underlying datasets date back to the 1980s and 1990s~\cite{McDonald:1986yq,Werlen:1988fh,Brandenburg:1994wf}, continuous efforts are ongoing to refine the theoretical framework and extraction methodology.  
In our comparison, shown in Fig.~\ref{fig:mpicomparison}, we include three representative modern analyses from the past five years~\cite{Barry:2021osv,Novikov:2020snp,Kotz:2025lio}.  
These phenomenological PDF sets are publicly available through the \texttt{LHAPDF6} library~\cite{Buckley:2014ana}, which provides a standardized interface and a consistent treatment of uncertainties.  

\texttt{LHAPDF} supports several uncertainty quantification schemes, two of which are used here.  
The first is the \textit{replica} method, in which experimental data are randomly shifted to produce an ensemble of equally probable replica PDFs, from which uncertainties are computed statistically.  
The second is the \textit{Hessian} method, where the uncertainty on an observable is represented by a set of $N$ eigenvector variations, allowing the reconstruction of symmetric or asymmetric errors from the corresponding Hessian matrix.  
All phenomenological PDFs are evaluated at the common energy scale $\mu=2~\text{GeV}$.  

The \texttt{JAM21PionPDFnlonll\_double\_Mellin} dataset produced by the JAM Collaboration~\cite{Barry:2021osv} includes deep-inelastic scattering (DIS) data from E615~\cite{McDonald:1986yq} and Drell-Yan data from NA10~\cite{Brandenburg:1994wf}, analyzed using NLO+NLL theory.  
PDFs are extracted through the double Mellin threshold resummation method~\cite{Ball:2007ra} and fitted to the functional form $x^\alpha (1-x)^\beta (1+\gamma x^2)$.  
The \texttt{FANTO10\_n15} dataset~\cite{Kotz:2025lio} also relies on E615 and NA10 data, but employs a broader uncertainty treatment that includes systematic effects from both the pion and the nuclear target PDFs.  
In this approach, Bezier curve fits are used to explore a less constrained functional space than traditional parametric forms, which results in noticeably larger error bands at low $x$.  
The \texttt{xFitterPI\_NLO} dataset~\cite{Novikov:2020snp} combines DIS data from E615, NA10, and photon production data from WA70~\cite{Werlen:1988fh}, analyzed within the \texttt{xFitter} framework.  
The PDFs are extracted using more flexible parametric forms than those adopted in the JAM analysis, leading to slightly larger final uncertainties.

In Fig.~\ref{fig:mpicomparison}, we also compare our results with other direct lattice QCD calculations of the connected contribution to the pion PDF moments, obtained using various lattice actions and pion masses.  
The calculations in Refs.~\cite{Alexandrou:2020gxs,Alexandrou:2021mmi} were performed in $N_f=2+1+1$ lattice QCD using boosted pion states in multiple directions to suppress power-divergent operator mixing.  
This approach, however, results in significantly noisier signals than in our calculation, despite employing statistics that are approximately six times larger for $n=3$ and one hundred times larger for $n=4$.
Although the pion mass used there ($m_\pi \simeq 260~\text{MeV}$) is lower than ours, the computation was carried out at a single lattice spacing, $a \simeq 0.093~\text{fm}$.  
Refs.~\cite{Gao:2022iex,Joo:2019bzr} employ the method of Ioffe-time distributions to extract PDF moments.  
In Ref.~\cite{Joo:2019bzr}, the calculation was performed at a single lattice spacing, $a=0.127~\text{fm}$, with a pion mass comparable to ours, while Ref.~\cite{Gao:2022iex} used one ensemble at the physical pion mass ($a=0.073~\text{fm}$) and two additional lattice spacings, $a=0.04$ and $0.06~\text{fm}$, at $m_\pi \simeq 300~\text{MeV}$.  

Overall, we observe good consistency between our results and recent phenomenological determinations, while a mild tension remains with the HadStruc results~\cite{Joo:2019bzr}.  
This discrepancy can be attributed to cutoff effects, since the HadStruc calculation was performed at a single, relatively coarse lattice spacing of $a\simeq0.127~\text{fm}$ without a continuum extrapolation.  
Our results agree well with phenomenological extractions and feature statistical uncertainties that are competitive with those of global fits.  
In the future, combining lattice QCD determinations such as ours with phenomenological analyses may further improve the precision of pion parton distribution functions, while still allowing for less constrained functional forms than those typically employed in purely parametric fits.

\section{Reconstruction of PDFs}
\label{sec:reconstruction}

The mathematical problem of reconstructing a function from a finite number of its moments has wide-ranging applications in many fields~\cite{JOHN20072890}.  
Modern approaches include adaptive spline algorithms~\cite{DESOUZA20102741}, kernel density estimators~\cite{ATHANASSOULIS2002273}, Gaussian processes, and entropy maximization methods~\cite{Biswas_2010,TAGLIANI1999291,Zhang:2023oja}.

For the pion PDFs, the simplest approach is to adopt the canonical ansatz~\cite{Sutton:1991ay}
\be
f(x) = \mathcal{N}\, x^{\alpha}(1-x)^{\beta} \, p(x)\,,
\label{eq:pdf_ansatz}
\ee
where, for each valence-quark distribution, the normalization $\mathcal{N}$ is fixed by
\be
\int_0^1 dx \, f_v(x) = 1\,.
\label{eq:pdf_norm}
\ee
In most of the literature, $p(x)$ is chosen as a low-order polynomial, sometimes supplemented by a $\sqrt{x}$ term or by Bernstein polynomials.  
We find that such extensions have no impact on our results and therefore set $p(x)=1$.  

For the moment ratios $\braket{x^m}/\braket{x}$, the corresponding analytic form reads
\be
\frac{\braket{x^m}}{\braket{x}} =
\frac{m+\alpha+\beta+2}{\alpha+\beta+3} 
\cdot 
\frac{\prod_{i=1}^{m-1} (\alpha+1+i)}{\prod_{i=1}^{m-1} (\alpha+\beta+3+i)}\,.
\label{eq:moment_ratio_param}
\ee
The parameters are constrained by $\alpha > -1$ and $\beta > -1$.  
The parameter $\beta$, which governs the large-$x$ behavior of the distribution, has been the subject of a long-standing debate in the literature.  
Most model predictions fall in the range $\beta = 1\text{--}2$~\cite{Ezawa:1974wm,Landshoff:1973pw,Gunion:1973ex,Farrar:1979aw,Berger:1979du,Shigetani:1993dx,Szczepaniak:1993uq,Davidson:1994uv,Hecht:2000xa,Melnitchouk:2002gh,Noguera:2015iia,Hutauruk:2016sug,Hobbs:2017xtq,deTeramond:2018ecg,Bednar:2018mtf,Lan:2019vui,Lan:2019rba,Chang:2020kjj,Cui:2020tdf,Kock:2020frx,Cui:2022bxn,Albino:2022gzs,Ahmady:2022dfv,Pasquini:2023aaf,Lu:2023yna}, 
while the most recent phenomenological extractions typically favor values of $\beta$ close to~1~\cite{Novikov:2020snp,Barry:2021osv}.
In our analysis, the parameters $\alpha$ and $\beta$ are determined by fitting the continuum-extrapolated lattice results for the ratios $\braket{x^m}/\braket{x}$ as a function of $m$, using Eq.~\eqref{eq:moment_ratio_param} as the fitting form.  
This strategy provides a way to reconstruct the shape of the pion PDF from the available nonperturbative moments.

Previous studies on lattice QCD data have noted that $\chi^2$ minimization with this functional form can be unstable and admit multiple minima~\cite{Joo:2019bzr}. In our dataset, even under the constraints $\alpha > -1$ and $\beta > -1$, we find that $\chi^2$ decreases to zero as $\alpha \to -1$. To stabilize the fit we impose physically motivated Bayesian priors on the parameters. Specifically, we use a normally distributed, $N(\mu,\sigma)$, prior $\beta \sim N(0,5)$, motivated by previous studies~\cite{Ezawa:1974wm,Landshoff:1973pw,Gunion:1973ex,Farrar:1979aw,Berger:1979du,Shigetani:1993dx,Szczepaniak:1993uq,Davidson:1994uv,Hecht:2000xa,Melnitchouk:2002gh,Noguera:2015iia,Hutauruk:2016sug,Hobbs:2017xtq,deTeramond:2018ecg,Bednar:2018mtf,Lan:2019vui,Lan:2019rba,Chang:2020kjj,Cui:2020tdf,Kock:2020frx,Cui:2022bxn,Albino:2022gzs,Ahmady:2022dfv,Pasquini:2023aaf,Lu:2023yna} which generally find $\beta < 5$. The precise choice of central value and width for this prior, provided it remains reasonably broad, has no impact on our final results, which give $\beta \sim 1$.

\begin{figure}
    \centering
    \includegraphics[width=\linewidth]{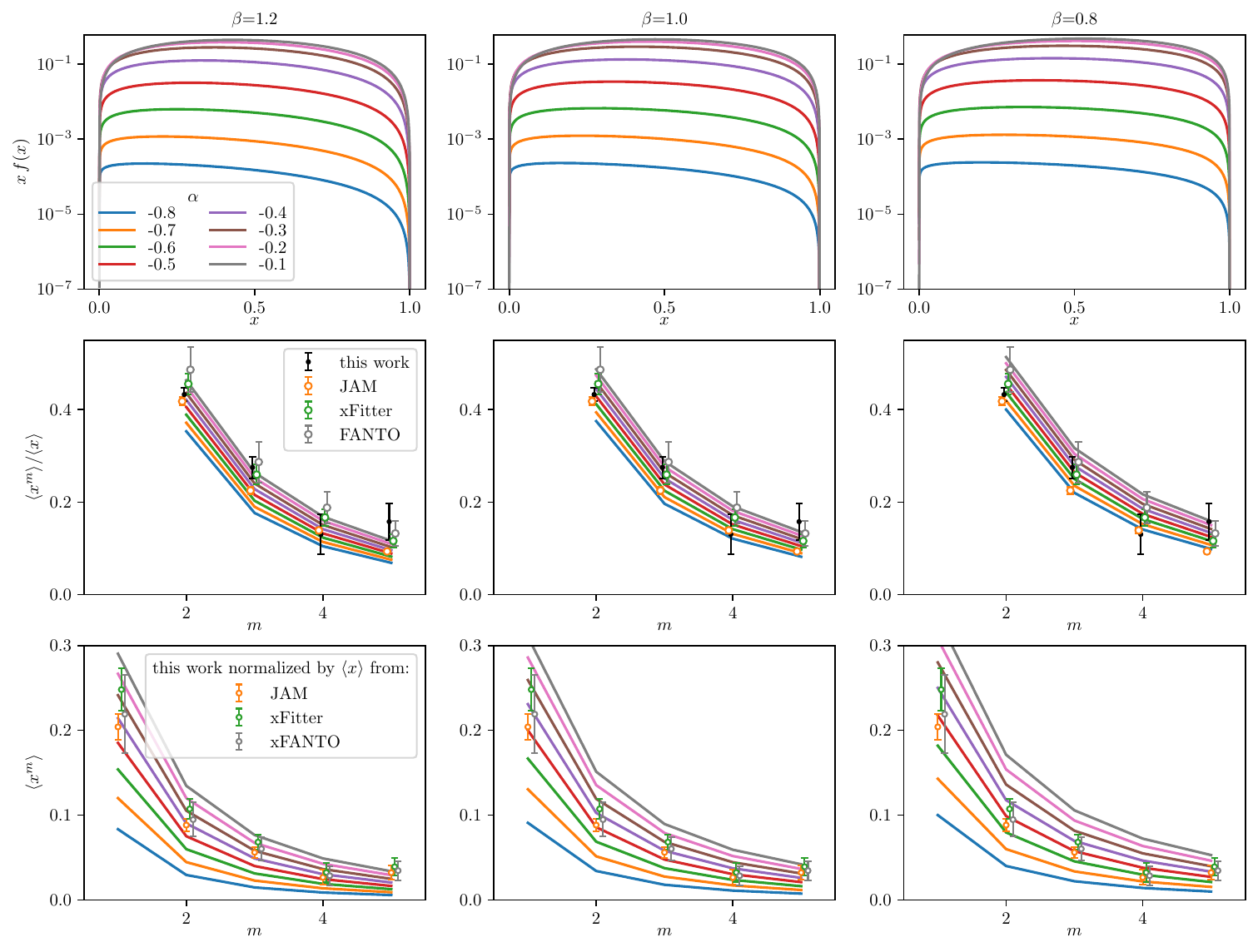}
    \caption{
Dependence of the pion valence PDF and its moments on the parameter $\alpha$ for three representative values of $\beta$ close to $\beta \sim 1$.  
The colored curves show $x\,f_v(x)$ and the corresponding moments, based on the parametrization in Eq.~\eqref{eq:pdf_ansatz}
and its implied expression for the ratios of moments
given in Eq.~\eqref{eq:moment_ratio_param}.  
\textbf{Top panels:} $x\,f_v(x)$ for eight values of $\alpha$ between $-0.8$ and $-0.1$.  
\textbf{Middle panels:} Ratios of moments $\braket{x^m}/\braket{x}$ compared with our lattice results (black points) and phenomenological extractions from JAM~\cite{Barry:2021osv}, xFitter~\cite{Novikov:2020snp}, and FANTO~\cite{Kotz:2025lio}.  
\textbf{Bottom panels:} Individual moments $\braket{x^m}$ obtained by combining our ratios with $\braket{x}$ values from the same three phenomenological analyses.  
The ratios show only a weak dependence on $\alpha$, whereas the reconstructed PDFs exhibit strong sensitivity—over the same $\alpha$ range.  
The lower moments display the highest sensitivity to $\alpha$.
}
    \label{fig:pdf_investigate}
\end{figure}

In contrast, the fit is highly sensitive to the prior on $\alpha$. Selecting it by minimizing the Gaussian Bayes Factor reproduces the same pathology as in the frequentist case, with the fit favoring values of $\alpha$ arbitrarily close to $-1$. To visualize the sensitivity to $\alpha$, in Fig.~\ref{fig:pdf_investigate} we plot $x\,f_v(x)$, $\braket{x^m}/\braket{x}$, and $\braket{x^m}$ for three values of $\beta$ close to $\beta \sim 1$, and eight values of $\alpha$ between $-0.8$ and $-0.1$. We observe that, for all $m$ considered, the moment ratios change only weakly with $\alpha$ (a small variation in the ratios spans a wide range of $\alpha$), implying that the ratios alone provide limited sensitivity to $\alpha$ unless the precision of the higher moments is significantly improved. By contrast, the PDF $x\,f_v(x)$ is highly sensitive to $\alpha$: over the same range of $\alpha$, the value of $x\,f_v(x)$ at its maximum decreases by several orders of magnitude.

Based on the physical prior that the value of $x\,f_v(x)$ at its maximum is predicted by the majority of phenomenological studies~\cite{Barry:2021osv,Novikov:2020snp,Kotz:2025lio} to be larger than $0.1$, we choose the prior $\log(\alpha+1) \sim N(-0.6,0.2)$. Decreasing the central value drives $\alpha \to -1$ and reduces $\chi^2$ towards zero, while increasing it leads to worse fits. Our choice yields reduced $\chi^2_\nu$ close to unity ($\sim 1.5$ including $n=6$, $\sim 1.2$ excluding it). If the width is increased by $0.3$ or more, the fit again pushes $\alpha \to -1$ and $\chi^2_\nu$ can be reduced arbitrarily. Although this specific choice of prior stabilizes the reconstruction, it is admittedly not ideal; in future work we will explore alternative strategies (e.g., Gaussian-process-based reconstructions) for extracting PDFs from ratios of moments.

The final results of this procedure yield $(\alpha,\beta)=(-0.48(10),\,0.93(16))$ when $n=6$ is included, and $(\alpha,\beta)=(-0.47(11),\,0.97(17))$ when it is not. The small impact of including $n=6$ is expected given its larger uncertainties, stemming from both the degraded signal-to-noise of higher-derivative operators (and additional trace-subtraction terms) and the more restricted safe flow-time window due to cutoff effects.

Figure~\ref{fig:pdf_investigate} also shows the individual moments $\braket{x^m}$, obtained by combining our ratios with $\braket{x}$ values from the three phenomenological extractions discussed above.  
The lower moments display the strongest sensitivity to $\alpha$, as expected, since $\alpha$ governs the small-$x$ behavior where these moments receive their dominant contributions.  
This observation highlights the importance of improving the precision of the lowest moments and motivates future direct calculations of $\braket{x}_v$, both using unflowed three-point functions with standard renormalization schemes (e.g., RI-MOM) and by employing flowed fermion fields with their corresponding renormalization.

\section{Conclusions and outlook}
\label{sec:conclusions}

In this work we have presented the first nonperturbative determination of pion PDF moment ratios $\braket{x^{n-1}}/\braket{x}$ up to $n=6$ using the gradient flow approach in lattice QCD.  
Our analysis, based on four lattice spacings and a nonperturbatively improved action, achieves full control over excited-state contamination, continuum extrapolation, and the limit to vanishing flow time $t \to 0$.

Within the present statistical accuracy, excited-state effects are negligible, and cutoff effects are well described by O($a^2$) scaling in regions of flow time larger than a minimal value of $t$ that increases with $n$, as expected.  
The continuum-extrapolated ratios, matched to the $\overline{\text{MS}}$ scheme at $\mu = 2~\text{GeV}$, display stable behavior across flow times, with only small residual $t$-dependence.  
The final results for the ratios exhibit a smooth trend with increasing $n$, consistent with phenomenological expectations for the pion valence distribution.

We have also made an exploratory attempt to reconstruct the valence parton distribution function of the pion from these ratios.  
While the reconstructed distribution is consistent with phenomenological analyses, we find the procedure to be numerically unstable, underscoring the need for more robust reconstruction strategies.  
Further studies will explore improved reconstruction strategies, such as Gaussian processes or alternative functional bases, see for example Ref.~\cite{Alexandrou:2020tqq,Candido:2024hjt,Dutrieux:2024rem,Medrano:2025cmg}, to achieve a greater stability and reduce potential bias.

These findings validate the method proposed in Ref.~\cite{Shindler:2023xpd} as a systematically improvable and computationally efficient approach to determining partonic structure from first principles.  
The analysis strategy developed here, combining flowed operators, improved contraction algorithms, and multi-ensemble continuum limit, provides a general template for future applications.

Future work will focus on extending this study in several directions.  
A natural next step is to repeat the present calculation at lighter pion masses, progressing toward the physical point, to investigate the pion-mass dependence of the flowed moment ratios and provide direct predictions for phenomenology. Given the discrepancies among existing lattice determinations of the pion-mass dependence, such studies are essential to establish a consistent picture and to quantify possible corrections toward the physical point. In the future, these moments could be directly incorporated into global PDF fits, together with experimental and lattice QCD data e.g.~\cite{JeffersonLabAngularMomentumJAM:2022aix,Barry:2025wjx}. 
We also plan to extend the analysis to higher moments $n>6$, where improving the control of the continuum limit (possibly through combined $(a,t)$ extrapolations) will be essential.  
Finally, the same methodology can be applied to the study of baryonic structure, in particular to determine the proton parton distribution functions and to explore the gluon and sea-quark contributions.  
Such extensions will enable a comprehensive, systematically improvable lattice determination of partonic structure directly from QCD.

Altogether, this study strengthens the bridge between lattice QCD and global parton analyses, bringing precision hadron-structure calculations closer to experimental phenomenology.

\begin{acknowledgements}
\textit{Acknowledgements}: 
We thank Robert Harlander, Martin Hoferichter, and Wally Melnitchouk for valuable discussions on the gradient flow in perturbation theory, the chiral dependence of PDF moments, and the reconstruction of PDFs from moments, respectively.
We thank the authors of Refs.~\cite{Gao:2022iex,Barry:2021osv} for providing their data. All phenomenological results in the comparison plots were obtained using the {\tt LHAPDF6} library~\cite{Buckley:2014ana}.
Numerical calculations for this work were performed using resources from the National Energy Research Scientific Computing Center (NERSC), a Department of Energy Office of Science User Facility, under NERSC award NP-ERCAP0027662; and the Gauss Centre for Supercomputing e.V. (www.gauss-centre.eu) on the GCS Supercomputer JUWELS~\cite{juelich2021juwels} at the Jülich Supercomputing Centre (JSC).
We also acknowledge the EuroHPC Joint Undertaking for awarding this project access to the EuroHPC supercomputer LEONARDO, hosted by CINECA (Italy) and LUMI at CSC (Finland).
The authors acknowledge support as well as computing and storage resources from GENCI on Adastra and Occigen (CINES), Jean-Zay (IDRIS) and Ir\`ene-Joliot-Curie (TGCC) under projects (2020-2024)-A0080511504 and (2020-2024)-A0080502271 as well as 2025-A0180516207.

A.F. acknowledges support by the National Science and Technology Council of Taiwan under grants 113-2112-M-A49-018-MY3 and 111-2112-M-A49-018-MY2. 
The work of R.K. is supported in part by the NSF Graduate Research Fellowship Program under Grant DGE-2146752. 
This research was supported by the National Research Foundation (NRF) funded
by the Korean government (MSIT)(No. RS-2025-02221606).
G.P. acknowledges funding by the Deutsche Forschungsgemeinschaft (DFG, German Research Foundation) - project number 460248186 (PUNCH4NFDI).
D.A.P is supported from the Office of Nuclear Physics, Department of Energy, under contract DE-SC0004658. 
A.S. acknowledges funding support from Deutsche Forschungsgemeinschaft (DFG, German Research Foundation) through grant 513989149 and under the National Science Foundation grant PHY-2209185. The work of A.W-L. was supported by the U.S. Department of Energy, Office of Science, Office of Nuclear Physics, under contract number DE-AC02-05CH11231.
The research of S.Z. is funded, in part, by l'Agence Nationale de la Recherche (ANR), project ANR-23-CE31-0019. For the purpose of open access, the author has applied a CC-BY public copyright licence to any Author Accepted Manuscript (AAM) version arising from this submission.
We acknowledge support from the DOE Topical Collaboration “Nuclear Theory for New Physics”, award No. DE-SC0023663. This manuscript was partially finalized at Aspen Center for Physics, which is supported by National Science Foundation grant PHY-2210452.

The {\tt OpenQCD}~\cite{openqcd}, {\tt Chroma}~\cite{Edwards:2004sx},  {\tt QUDA}~\cite{Clark:2009wm,Babich:2011np,Clark:2016rdz}, {\tt QDP-JIT}~\cite{Winter:2014dka}, and {\tt LALIBE}~\cite{lalibe} software libraries were used in this work. 
Data analysis used {\tt NumPy}~\cite{harris2020array}, {\tt SciPy}~\cite{2020SciPy-NMeth}, {\tt gvar}~\cite{peter_lepage_2020_4290884}, and {\tt lsqfit}~\cite{peter_lepage_2020_4037174}.
Figures were produced using {\tt matplotlib}~\cite{Hunter:2007}.
\end{acknowledgements}

\bibliography{main}

\end{document}